%% file: HPCA2024_oTransformer.tex
\documentclass[10pt,conference]{IEEEtran}
\usepackage{mathptmx}

\usepackage{fancyhdr}
\usepackage[normalem]{ulem}
\usepackage[hyphens]{url}
\usepackage[sort,nocompress]{cite}
\usepackage{textcomp}

\usepackage{algorithm}
\usepackage{graphicx}
\usepackage{subfig}
\usepackage{threeparttable}
\usepackage{color}
\usepackage{colortbl}
\usepackage{multirow}
\usepackage{amsmath}
\usepackage{amsfonts}
\usepackage{comment}
\usepackage{bm}
\usepackage{etoolbox}                                      %
\usepackage{url}
\usepackage{nth}
\usepackage{balance}
\usepackage{courier}                                       %
\usepackage{booktabs}                                      %
\usepackage{listings}             

\usepackage[]{algpseudocode}
\usepackage{xspace}
\usepackage{xcolor}
\input{macro}
\usepackage{pifont}

\usepackage[final]{microtype}
\usepackage{flushend}
\usepackage[bookmarks=true,breaklinks=true,colorlinks,citecolor=citecolor,linkcolor=magenta,urlcolor=blue]{hyperref}

\newcommand{\hpcayear}{2024}

\title{Lightening-Transformer: A Dynamically-operated Optically-interconnected  \\ 
Photonic Transformer Accelerator}

\def\hpcacameraready{} %

\newcommand\hpcaauthors{
 Hanqing Zhu\textsuperscript{1},
Jiaqi Gu\textsuperscript{1,3},
Hanrui Wang\textsuperscript{2},
Zixuan Jiang\textsuperscript{1}, \\
Zhekai Zhang\textsuperscript{2},
Rongxing Tang\textsuperscript{1},
Chenghao Feng\textsuperscript{1}, \\
Song Han\textsuperscript{2},
Ray T. Chen\textsuperscript{1},
David Z. Pan\textsuperscript{1}
}
\newcommand\hpcaaffiliation{
\textsuperscript{1}The University of Texas at Austin, \textsuperscript{2}Massachusetts Institute of Technology, \textsuperscript{3}Arizona State University
}
\newcommand\hpcaemail{
\textit{\{hqzhu, jqgu, zixuan, rt970117, fengchenghao1996\}@utexas.edu, \{hanrui, zhangzk, songhan\}@mit.edu}, \\
\textit{\{chen, dpan\}@ece.utexas.edu}
}

\def\aeopen{}           %
\def\aereviewed{}     %
\def\aereproduced{} %

\input{hpca-template}

\begin{document}
\maketitle

\ifdefined\hpcacameraready 
  \thispagestyle{camerareadyfirstpage}
  \pagestyle{empty}
\else
  \thispagestyle{plain}
  \pagestyle{plain}
\fi

\newcommand{\hpcaheight}{0mm}
\ifdefined\eaopen
\renewcommand{\hpcaheight}{12mm}
\fi

\input{doc/0abstract}

\input{doc/1intro}

\input{doc/2prelim}

\input{doc/4algo}

\input{doc/arch}

\input{doc/6result}
\input{doc/discuss}

\input{doc/7conclu}

\input{doc/8acknowledge}

\newpage
\input{doc/appendix}

\newpage
\bibliographystyle{IEEEtranS}
\bibliography{./ref/Top_sim,./ref/NN,./ref/NP,./ref/ALG, ./ref/HWA}

\end{document}

%% file: macro.tex
\definecolor{citecolor}{RGB}{34,139,34}
\definecolor{mydarkblue}{rgb}{0,0.08,1}
\definecolor{mydarkgreen}{rgb}{0.02,0.6,0.02}
\definecolor{mydarkred}{rgb}{0.8,0.02,0.02}
\definecolor{mydarkorange}{rgb}{0.40,0.2,0.02}
\definecolor{mypurple}{RGB}{111,0,255}
\definecolor{myred}{rgb}{1.0,0.0,0.0}
\definecolor{mygold}{rgb}{0.75,0.6,0.12}
\definecolor{myblue}{rgb}{0,0.2,0.8}
\definecolor{mydarkgray}{rgb}{0.,0.2,0.2}

\definecolor{lightred}{RGB}{255,235,235}
\definecolor{lightgreen}{RGB}{235,255,235}
\definecolor{lightblue}{RGB}{235,235,255}
\definecolor{lightcyan}{RGB}{235,255,255}
\definecolor{lightmagenta}{RGB}{255,235,255}
\definecolor{lightyellow}{RGB}{255,255,235}

\definecolor{qxkcolor}{RGB}{215,235,255}
\definecolor{softmaxcolor}{RGB}{230,235,255}
\definecolor{probxvcolor}{RGB}{255,255,235}

\definecolor{topkcolor}{RGB}{255,235,235}
\definecolor{zecolor}{RGB}{255,255,235}
\definecolor{dynacolor}{RGB}{235,255,255}

\definecolor{reviewcolor}{RGB}{0,0,200}

\definecolor{mod1color}{RGB}{192, 0, 0}
\definecolor{DAC1color}{RGB}{177, 102, 255}
\definecolor{mod2color}{RGB}{221, 133, 131}
\definecolor{DAC2color}{RGB}{0,174,186}
\definecolor{lasercolor}{RGB}{248, 182, 45}
\definecolor{ADCcolor}{RGB}{132, 60, 12}
\definecolor{detcolor}{RGB}{112, 173, 71}

\algdef{SE}[DOWHILE]{Do}{doWhile}{\algorithmicdo}[1]{\algorithmicwhile\ #1}%

\newcommand{\name}{\texttt{Lightening-Transformer}\xspace}
\newcommand{\sname}{\texttt{LT}\xspace}

\newcommand{\oDot}{\texttt{DDot}\xspace}
\newcommand{\oMM}{\texttt{DPTC}\xspace}

\newcounter{rlabelno}

\usepackage{soul}

%% file: hpca-template.tex
\author{
  \ifdefined\hpcacameraready
    \IEEEauthorblockN{\hpcaauthors{}}
      \IEEEauthorblockA{
        \hpcaaffiliation{} \\
        \hpcaemail{}
      }
  \else
    \IEEEauthorblockN{\normalsize{HPCA \hpcayear{} Submission
      \textbf{\#\hpcasubmissionnumber{}}} \\
      \IEEEauthorblockA{
        Confidential Draft \\
        Do NOT Distribute!!
      }
    }
  \fi 
}

\fancypagestyle{camerareadyfirstpage}{%
  \fancyhead{}
  
  \fancyhead[C]{
    \ifdefined\aeopen
    \parbox[][12mm][t]{13.5cm}{\hpcayear{} IEEE International Symposium on High-Performance Computer Architecture (HPCA)}    
    \else
      \ifdefined\aereviewed
      \parbox[][12mm][t]{13.5cm}{\hpcayear{} IEEE International Symposium on High-Performance Computer Architecture (HPCA)}
      \else
      \ifdefined\aereproduced
      \parbox[][12mm][t]{13.5cm}{\hpcayear{} IEEE International Symposium on High-Performance Computer Architecture (HPCA)}
      \else
      \parbox[][0mm][t]{13.5cm}{\hpcayear{} IEEE International Symposium on High-Performance Computer Architecture (HPCA)}
    \fi 
    \fi 
    \fi 
    \ifdefined\aeopen 
      \includegraphics[width=12mm,height=12mm]{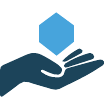}
    \fi 
    \ifdefined\aereviewed
      \includegraphics[width=12mm,height=12mm]{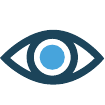}
    \fi 
    \ifdefined\aereproduced
      \includegraphics[width=12mm,height=12mm]{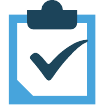}
    \fi
  }
  \fancyfoot[C]{}
}

%% file: doc/0abstract.tex
\begin{abstract}
\label{abstract}
The wide adoption and significant computing resource cost of attention-based transformers, e.g., Vision Transformers and large language models, have driven the demand for efficient hardware accelerators.
While electronic accelerators have been commonly used, there is a growing interest in exploring photonics as an alternative technology due to its high energy efficiency and ultra-fast processing speed.
Photonic accelerators have demonstrated promising results for convolutional neural networks (CNNs) workloads, which predominantly rely on weight-static linear operations. 
However, they encounter challenges when it comes to efficiently supporting attention-based Transformer architectures, raising questions about the applicability of photonics to advanced machine-learning tasks.
The primary hurdle lies in their inefficiency in handling the unique workloads inherent to Transformers, i.e., dynamic and full-range tensor multiplication.

In this work, we propose \name, the \textit{first} light-empowered, high-performance, and energy-efficient photonic Transformer accelerator.
To overcome the fundamental limitation of existing photonic tensor core designs, we introduce a novel dynamically-operated photonic tensor core, \oMM, consisting of a crossbar array of interference-based optical vector dot-product engines, supporting highly parallel, dynamic, and full-range matrix multiplication.
Furthermore, we design a dedicated accelerator that integrates our novel photonic computing cores with photonic interconnects for inter-core data broadcast, fully unleashing the power of optics.
The comprehensive evaluation demonstrates that \name achieves $>$$2.6\times$ energy and $>$$12\times$ latency reductions compared to prior photonic accelerators and delivers the lowest energy cost and 2 to 3 orders of magnitude lower energy-delay product compared to the electronic Transformer accelerator, all while maintaining digital-comparable accuracy.
Our work highlights the immense potential of photonics for efficient hardware accelerators, particularly for advanced machine-learning workloads, such as Transformer-backboned large language models (LLM).
Our implementation is available at \href{https://github.com/zhuhanqing/Lightening-Transformer}{https://github.com/zhuhanqing/Lightening-Transformer}.

\end{abstract}

%% file: doc/1intro.tex
\section{Introduction}
\label{sec:Introduction}

Recently, attention-based Transformers have gained immense popularity and demonstrated remarkable success in various domains, e.g., natural language processing (NLP) \cite{NN_NAACL2019_Kenton, NN_Neurips2020_Brown, openai2023gpt4, xiao2023streamingllm} and computer vision (CV)~\cite{NN_ICLR2021_Alexey, NN_ECCV2020_Carion, NN_ICML2021_Touvron}. The attention mechanism enables dynamic feature aggregation, long-distance modeling, and global context extraction, contributing significantly to the impressive performance \cite{NN_NIPS2017_vaswani, NN_ICLR2021_Alexey}.
However, this exceptional performance comes at a considerable computational cost. The quadratic complexity of attention, in terms of computation and memory, combined with a large number of parameters, demands substantial computational resources.
This poses a challenge for deploying Transformers, particularly in resource-constrained systems where such computational demands are prohibitive.
Hence, there is a pressing need to develop domain-specific hardware accelerators for the efficient deployment of Transformers in real-world applications.

\begin{figure}
    \centering
    \includegraphics[width=0.45\textwidth]{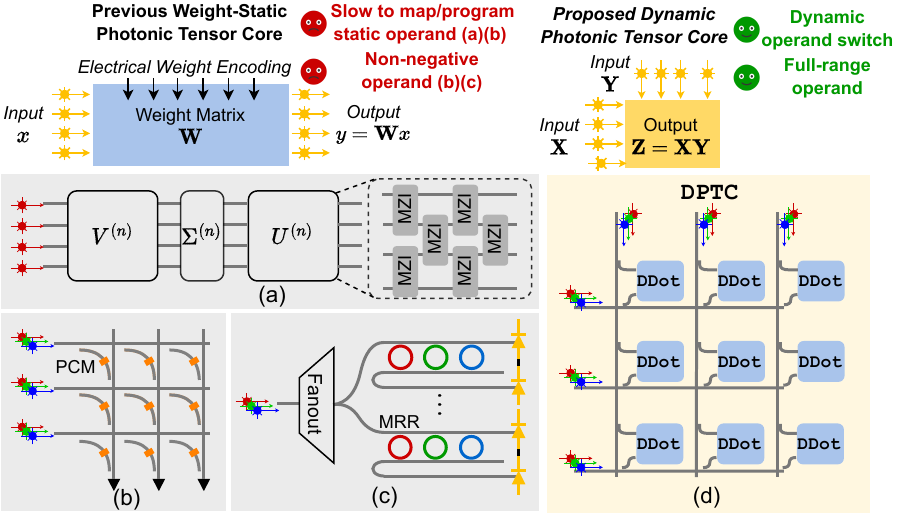}
    \small{\caption{(a), (b), (c) Prior weight-static photonic tensor core designs~\cite{NP_NATURE2017_Shen, NP_SciRep2017_Tait, NP_Nature2021_Feldmann}.
    (d) Our proposed dynamic photonic tensor core design without static weight constraints.}
    \label{fig:prior_onn}}
    \vspace{-15pt}
\end{figure}

Several hardware accelerators based on digital electronics have been proposed to accelerate the inference of Transformers~\cite{HWA_arxiv2022_Sun, HWA_HPCA2021_Wang, HWA_HPCA2022_Zhou, HWA_HPCA2023_Dong, HWA_HPCA2023_You}.
However, traditional electrical digital computing platforms face significant challenges as transistor-based chips reach the limits of Dennard scaling, leading to increased power dissipation per unit area and diminishing performance improvements. 
As a compelling alternative, integrated photonic accelerators~\cite{NP_NaturePhotonics2021_Shastri, NP_TCAS2022_Gu} have emerged as next-generation computation platforms offering ultra-high speed, high parallelism, and low energy consumption.
Various optical systems are actively being studied for accelerating convolutional neural network (CNN) workloads with different photonic tensor core (PTC) designs, e.g., Mach-Zehnder interferometer~(MZI) array~\cite{NP_NATURE2017_Shen}, Micro-ring Resonator~(MRR) bank~\cite{NP_SciRep2017_Tait, NP_DAC2021_Sunny}, and non-volatile Phase Change Material~(PCM)-based crossbar~\cite{NP_Nature2021_Feldmann}. 

However, existing photonic accelerators are mainly designed and optimized for weight-static NNs, e.g., CNNs, where convolution and fully-connected layers only involve matrix multiplications~(MM) between the learned static weight matrix and (non-negative) input tensors.
They fail to efficiently support attention-based Transformers due to the following \textit{challenges}:

\textbf{Matrix multiplication with two dynamic input operands.}
Unlike digital electronics, prior photonic accelerators typically need to map one operand of MM (usually the weight matrix $\mathbf{W}$) onto PTC's circuit transmission by programming its devices, shown in Figure~\ref{fig:prior_onn}.
This procedure typically involves costly operand mapping and slow device programming, leading to a preference for keeping the encoded operand \textit{static} and reusing it for many inputs to amortize the cost, i.e., ``weight-static''.
For example, MZI array~\cite{NP_NATURE2017_Shen} requires singular value decomposition (SVD) and phase decomposition for operand mapping.
To map a 12$\times$12 matrix, it takes $\sim$1.5~ms for SVD and phase decomposition on CPU.
Moreover, facing the challenges of bulky area and large insertion loss, 
PTC designs prefer to use low-loss, compact, and non-volatile devices that usually cost 10~ns-10~$\mu$s to be programmed~\cite{NP_arxiv2023_anderson, NP_MN2023_Quack, NP_Nature2021_Feldmann}.
However, 
attention in Transformers is built on dynamic matrix multiplication, with both operands being dynamically generated activations.
This dynamic nature necessitates frequent real-time operand mapping and device reprogramming.
Real-time mapping and reprogramming are generally not affordable or challenging to amortize given the orders-of-magnitude higher runtime than ultra-fast computing ($<$100~ps) and limited reuse opportunity in the dynamic MM scenario, incurring long latency for preparing PTCs and leading to severe system stall.

\textbf{Matrix multiplication with full-range input operands.}
Transformers require full-range matrix multiplications as activations are not constrained to non-negative only.
However, 
prior incoherent PTCs, such as MRR bank, pose range limitations on operands.
At least one of the operands is limited to be non-negative as their computation is based on light intensity modulation (non-negative only).
The absence of full range support often requires decomposing full-range operands into differences of non-negative operands $(X_+-X_-)(Y_+-Y_-)$ and processing $X_+Y_+$, $X_+Y_-$, $X_-Y_+$, and $X_-Y_-$ separately~\cite{NP_DAC2021_Sunny, NP_arxiv2023_anderson} with extra accumulation steps, incurring $>$$2$-$4\times$ hardware cost compared to one with full-range operand support.
Full-range operand encoding is a unique feature of coherent PTCs where signs can be encoded to phases and processed via interference.
Coherent MZI array indeed can achieve full-range MVM with a single wavelength. 
In comparison, our PTC design maintains the full-range feature with a novel interference circuit design while utilizing multiple wavelengths for ultra-parallel MM.

Above all, existing photonic accelerators encounter significant difficulties in efficiently accelerating Transformers' unique dynamic and full-range MM workloads.
To address those challenges, 
in this work,
we propose the \textit{first} customized photonic accelerator to support attention-based Transformers,
named \name.
We first design a \textit{coherent} dot-product unit \oDot that enables multiplication between two dynamically encoded full-range optical vectors.
\oDot directly encodes both operands as high-speed \emph{coherent} optical signals, 
thus can represent signs as signal phases and support real-time operand switching with negligible mapping or programming cost ($<$10~ps).
The dot-product mechanism is based on coherent light interference with a coupler and balanced photodetection, further enabling to detect full-range outputs as positive or negative photocurrent. 
Based on \oDot,
we devise a novel crossbar-style photonic tensor core, \oMM, for ultra-parallel and energy-efficient dynamic full-range matrix multiplication, which is the key building block of our Transformer accelerator.
We fully leverage both spectral and spatial parallelism of optics by exploiting the wavelength-division multiplexing (WDM) capacity,
and unleash the natural optical broadcast capability to enable intra-core and inter-core operand sharing.
As a result, we can maximize the hardware sharing, trim modulation overhead, and significantly boost processing parallelism and efficiency.

The major contributions of this paper are as follows:
\begin{itemize}
    \item We present, \textit{for the first time}, a light-empowered, high-performance, and energy-efficient photonic Transformer accelerator, dubbed  \name, overcoming the efficiency and flexibility limitations of prior photonic accelerators for Transformer acceleration.
    \item We design a novel dynamically-operated photonic tensor core, \oMM, that enables ultra-parallel and energy-efficient dynamic full-range general matrix multiplication in a one-shot way. 
    \oMM utilizes the WDM technique to enable spectral parallelism and a crossbar-based circuit topology to explore intra-core operand sharing, offering exceptional processing parallelism and energy efficiency.
    \item We develop a dedicated dynamically-operated optically-interconnected accelerator that utilizes our superior photonic tensor cores for efficient computing and optical interconnects for efficient inter-core data broadcast to fully unleash the power of optics.
    To further reduce the signal conversion cost, we employ architecture-level optimization to reduce input electrical-to-optical (E-O) conversion cost via sharing optical signals inter-core and reduce output optical-to-electrical (O-E) conversion cost via analog-domain temporal accumulation.
    \item We evaluate proposed \name comprehensively compared to photonic and electronic accelerators across different Transformer benchmarks.
    \name significantly outperforms prior photonic designs, achieving over 2.6$\times$ energy and over 12$\times$ latency reductions.
    Furthermore, compared to state-of-the-art electronic Transformer accelerators, our accelerator consistently delivers the lowest energy consumption and achieves 2 to 3 orders of magnitude lower energy-delay product while keeping digital-comparable accuracy. 
\end{itemize}

%% file: doc/2prelim.tex
\section{Preliminaries and Background}
\label{sec:Background}

\input{tablels/arch_compare}

\subsection{Transformer and Self-Attention}
Transformer~\cite{NN_NIPS2017_vaswani} was initially proposed as a sequence transduction model for NLP tasks.
Three mainstream Transformer architectures are encoder-decoder (BERT \cite{NN_NAACL2019_Kenton}, ViT\cite{NN_ICLR2021_Alexey}), causal decoder (GPT-series \cite{gpt3}), and the prefix decoder (GLM-130B \cite{glm-130b}).
Despite different Transformer architectures, they are usually a stack of several identical blocks.
Both encoder and decoder blocks comprise a multi-head self-attention~(MHA) module, a feed-forward network~(FFN), the shortcut connection, and layer normalization (LN)~\cite{layernorm}.
The decoder additionally adds cross-attention and masked self-attention modules.
We use the basic encoder block as an example, defined as 
\begin{equation}
    \begin{aligned}
        \mathbf{X}^{'}_{l+1} &= \text{MHA}(\text{LN}(\mathbf{X}_{l})) + \mathbf{X}_{l}; \\
        \mathbf{X}_{l+1} &= \text{FFN}(\text{LN}(\mathbf{X}_{l+1}^{'})) + \mathbf{X}^{'}_{l+1},
    \end{aligned}
\end{equation}
where $\mathbf{X}_{l}$ is input sequences of $l$-th layer.

\noindent\textbf{Multi-head Self-Attention (MHA)}.
Attention is a novel feature of Transformers where pairwise correlations across the entire input sequence are computed.
MHA has $H$ self-attention heads.
In each head, the input vector is transformed into the query ($\mathbf{Q}$), key ($\mathbf{K}$), and value ($\mathbf{V}$) vectors by linear projection. 
Then, 
the attention function between different input vectors is
\begin{equation}
    \text{Attention}(\mathbf{Q}, \mathbf{K}, \mathbf{V}) = \text{softmax} \big(\mathbf{Q} \mathbf{K}^T/\sqrt{d_k}\big) \mathbf{V},
\end{equation}
where $d_k$ is $\mathbf{Q}$ and $\mathbf{K}$'s dimension.
Although the attention is still based on matrix multiplication,
notably, it is totally different from the linear layer, which poses unique challenges for photonic inference accelerators.
In a linear layer, the MM is conducted between the static weight matrix and the dynamic input matrix.
In contrast, attention requires the MM between dynamically generated matrices, i.e., $\mathbf{Q}$, $\mathbf{K}$, and $\mathbf{V}$.

\noindent\textbf{Feed-forward Network (FFN)}.
The FFN module usually contains two linear layers with an activation function in between. 
GELU~\cite{gelu} is a popular option with better performance.

\subsection{Optical Computing Device Basics}
\label{subsec:basics}

\noindent\textbf{Phase shifter (PS)}: PS is an active device that produces a controlled phase shift $\phi$ on the light signal $x$ by manipulating the waveguide's effective refractive index.
The output is $e^{j\phi} x$.

\noindent\textbf{Directional coupler (DC)}: DC is a passive device that can produce interference between two coherent light signals.
The device consists of two waveguides positioned close to each other such that energy can transfer between them.
Its transfer matrix of a 2-by-2 DC is
$\scriptsize\begin{bmatrix}
t             & \sqrt{1-t^2}j \\
\sqrt{1-t^2}j & t 
\end{bmatrix}$,
where $t$ is the transmission coefficient. $t$=$\sqrt{2}/2$ in a 3 dB 50:50 DC.

\noindent\textbf{Mach-Zehnder modulator (MZM)}: MZM starts with one splitter that splits the optical input $E_{in}$ into upper and lower modulator arms.
After being phase modulated, the signals from the two arms are recombined as the optical output $E_{out}$.
With equal splitting and differential phase shift, $+\phi$ and $-\phi$ on the two arms~\cite{NP_JLT1988_Koyama},
respectively, full-range encoding $E_{out} = E_{in} \cos \phi$ can be achieved with MZM by tuning $\phi \in [0, \pi]$.

\noindent\textbf{Microring/microdisk resonator}: Micro-ring (MRR) and microdisk (MD) resonators are compact photonic devices that serve as narrowband filters to enable the transmission of a certain wavelength.
They can be utilized for constructing optical switches and WDM MUX and DEMUX units.

\noindent\textbf{Mach-Zehnder interferometer (MZI)}: MZI consists of two cascaded directional couplers and two phase shifters.
It can perform arbitrary 2-D unitary matrix operations, thus serving as a basic building block for constructing the MZI array~\cite{NP_NATURE2017_Shen}.

\subsection{Challenges of Prior Photonic Accelerators for Transformer}
\label{subsec:challenge}
\noindent\textbf{Unique features of Transformer workloads.}
Compared to weight-static NN architectures, the unique execution patterns of Transformers have brought unique workload characteristics.
\ding{202}
Attention modules require \textbf{MM with two dynamic input operands}, which requires frequent operand switching and real-time operand mapping and device programming.
If operand mapping and device programming are fairly slow compared to ultra-fast computing speed, it will result in severe system stall and significantly increase latency.
To clarify, we define the concepts of \emph{dynamic} and \emph{static} operands. 
In attention, both operands are input-dependent activations generated in real-time, rendering them dynamic. 
In contrast, weight matrices are fixed (static) during the inference.
\ding{203} Transformer requires \textbf{MM with full-range operands}, as activations are not restricted to non-negative only. 
This is especially true across Transformer layers with the widespread utilization of GELU and LayerNorm.

\noindent\textbf{Prior photonic accelerators.}
In this paper, we focus on universal linear units that can potentially support MHA and FFN in Transformers. Specifically, we consider MZI array~\cite{NP_NATURE2017_Shen}, MRR bank~\cite{NP_SciRep2017_Tait, NP_DAC2021_Sunny}, and non-volatile PCM-based crossbar~\cite{NP_Nature2021_Feldmann}. We omit the discussion of sub-space or convolution-specialized ones~\cite{NP_ASPDAC2020_Gu, NP_ACS2022_Feng, NP_NatureComm_Zhu, NP_ISCA2021_shiflett, li2023photofourier}.
Table~\ref{tab:compare_onns} provides a comprehensive comparison of operand characteristics, mapping \& programming cost, operation type (MM/MVM) between PTC designs of existing accelerators and our PTC design, \oMM. 
We highlight whether they can efficiently support unique Transformer workloads, i.e., dynamic MM and full-range MM.

\noindent\textbf{Challenge 1: Prior weight-static PTCs fail to efficiently support dynamic MMs}.
Before performing optical computing, we need to map operands onto PTC and program devices to get the desired transfer matrix.
However, MZI array requires extra SVD and phase decomposition to obtain phase information needed for device programming. 
As both operands of attention are activations, 
this complicated operand mapping step needs to be performed at runtime, which can introduce significant delay.
For instance, on a CPU, the required SVD and phase decomposition step takes $\sim$1.5 ms for a 12$\times$12 matrix.
As a result, severe system stalls occur, making the use of MZI array impractical for dynamic MM scenarios.
Besides, current PTC designs face challenges in terms of bulky area and huge insertion loss. Thus, they favor low-loss, compact, and non-volatile devices, as in the MZI array and PCM crossbar. 
However, these devices suffer from slow programming(10~ns-10~$\mu$s) that cannot be easily amortized, given
the orders-of-magnitude higher programming latency than ultra-fast computing.

\noindent\emph{\underline{Insight 1}: Mapping and device programming can dominate total latency for weight-static PTCs.}
Given the huge gap between ultra-fast computing speed and slow mapping/reprogramming speed, the overhead of setting up weight-static cores can hardly be amortized across batch/token dimensions.
To eliminate this dominant mapping cost and largely reduce the total latency, we can optically encode both operands for fast dynamic switching.
\noindent\textbf{Challenge 2: Prior incoherent PTCs fail to efficiently support full-range MMs}.
As an incoherent design, at least one of the operands of MRR bank is limited to be non-negative as its computation is
based on light intensity modulation.
The absence of full-range support requires decomposing full-range operands into differences of non-negative ones $(X_+-X_-)(Y_+-Y_-)$~and processing $X_+Y_+$, $X_+Y_-$, $X_-Y_+$, and $X_-Y_-$ separately by~multiple inferences or duplicated PTCs~\cite{NP_DAC2021_Sunny, NP_ISCA2021_shiflett}, incurring $>$$2$-$4\times$ extra hardware cost compared to one with full-range~support.
Since the modulation/DAC cost of inputs is doubled,
this will eliminate the advantages gained from amortizing DAC and dynamic modulation costs associated with weight-static dataflow.

\noindent\emph{\underline{Insight 2}: Phases of light can boost information processing throughput}.~
Instead of performing non-negative matrix multiplication using the weighted sum of light intensities like incoherent PTCs,
coherent PTC allows more information to propagate through the circuit by encoding signs to the extra phase dimension and performing signed computation via interference.
Hence, we can design a coherent PTC to efficiently support one-shot full-range MM without the above decomposition overhead.

%% file: tablels/arch_compare.tex
\begin{table*}[ht]
\centering
\small{
\caption{
Comparison of our dynamically-operated photonic tensor core, \oMM, to prior PTC designs.
Each PTC takes two operands to perform either matrix-vector multiplication (MVM) or matrix multiplication (MM).
Previous PTC designs fail to efficiently support (1) dynamic MM in attention due to the high cost of operand mapping or device programming and (2) full-range MM with no extra overhead, such as duplicated PTCs or multiple inferences.
}
\label{tab:compare_onns}
}
\resizebox{0.95\textwidth}{!}{
\begin{tabular}{c|c|c|c|c|c}
\hline
PTC Designs & MZI array~\cite{NP_NATURE2017_Shen}   & PCM Crossbar~\cite{NP_Nature2021_Feldmann}          & MRR Bank 1~\cite{NP_SciRep2017_Tait}              & MRR Bank 2~\cite{NP_DAC2021_Sunny}              & Ours \oMM \\ \hline
Operand 1                       & Static, Full-range   & Static, Positive Only   & Dynamic, Full-range    & Dynamic, Positive Only  & Dynamic, Full-range                               \\
Operand 2                         & Dynamic, Full-range  & Dynamic, Positive Only  & Dynamic, Positive Only & Dynamic, Positive Only  & Dynamic, Full-range                               \\
Mapping \& Programming Cost                   & High     & Medium            & Low                 & Low                                  & \cellcolor[HTML]{FFCE93}Low                       \\
Operation Type             & MVM     & MM              & MVM                    & MVM                                       & \cellcolor[HTML]{FFCE93}MM  \\
Dynamic MM Support (Attention)             & \ding{56}                 & \ding{56}                    & \ding{52}                     & \ding{52}                     & \cellcolor[HTML]{FFCE93} \ding{52}                     \\
Full-range MM Support (No overhead)             & \ding{52}                 & \ding{56}                    & \ding{56}                     & \ding{56}                     & \cellcolor[HTML]{FFCE93} \ding{52}                     \\
\hline
\end{tabular}
}
\vspace{-10pt}
\end{table*}

%% file: doc/4algo.tex
\section{Proposed Photonic Tensor Core Design}
To address the above-mentioned challenges of previous designs, we clarify the ultimate goal of this work and introduce our novel PTC design following the above important insights.

\noindent\textbf{Goal: a customized photonic tensor core design for efficient Transformer acceleration} with three essential traits:
\begin{itemize}
    \item Support for \textbf{general matrix multiplications}.
    \item Support for \textbf{full-range inputs/outputs}.
    \item Efficient handling of \textbf{dynamic operands with low encoding and signal modulation cost}.
\end{itemize}

To meet these traits, we first propose a dynamically-operated dot-product engine, \oDot, capable of computing the dot-product of two \textit{dynamically-encoded} \textit{full-range} optical vectors. 
Based on the basic building block \oDot, we then introduce a crossbar-style PTC, \oMM, with maximized intra-core operand sharing, enabling ultra-parallel and energy-efficient MM.

\subsection{DDot: Dynamically-Operated Full-range Dot-Product Unit}
To perform optical dot-product between vectors $\mathbf{x}$ and $\mathbf{y}$, we design a \oDot based on coherent interference shown in Figure~\ref{fig:photonic_arch}(a).
We employ the wavelength-division multiplexing (WDM) technique and encode each input pairs $(x_i, y_i)$ in the same wavelength $\lambda_i$.
We input the WDM light signals carrying $\mathbf{x}$ and $\mathbf{y}$ through the two arms of $50:50$ directional coupler (DC) with a $-90^{\circ}$ phase shifter (PS) on DC's upper left port.
Considering each input pair $(x_i, y_i)$ at the same wavelength,
the outputs at the right and left output ports of the DC, $z_i^0$ and $z_i^1$ can be computed as,
\begin{equation}
    \label{eq:signal_pair}
    \begin{aligned}
       \begin{pmatrix}
            z_i^0\\
            z_i^1
        \end{pmatrix}
       &=\frac{1}{\sqrt{2}} \begin{pmatrix}
            1 & j\\
            j & 1
        \end{pmatrix} 
        \begin{pmatrix}
            1 & 0\\
            0 & e^{-j\frac{\pi}{2}}
        \end{pmatrix}
        \begin{pmatrix}
            x_i \\
            y_i
        \end{pmatrix} \\
        &=\frac{1}{\sqrt{2}}\begin{pmatrix}
            x_i + y_i\\
            j(x_i -y_i)
        \end{pmatrix}.
    \end{aligned}
\end{equation}
The two output signals are orthogonal in the complex plane.
By adopting broadband devices,
we can have the same transfer function for a range of wavelengths.
In this way, with WDM signals, 
each wavelength with $(x_i, y_i)$ encoded interferes in parallel following Eq. \eqref{eq:signal_pair}, while different wavelengths don't interfere.
The photo-diode (PD) at the end of each output port of DC converts the incident WDM signals to the photocurrent.
The generated photocurrent is proportional to the accumulated intensities of the WDM signals, which is the square of optical magnitudes.
Thus, the photocurrents generated at the right and left PD denoted as $I^0$ and $I^1$, can be expressed as,
\begin{equation}
    \label{eq:power_pair}
    \begin{aligned}
    \begin{pmatrix}
            I^0\\
            I^1
        \end{pmatrix} 
        & \propto\frac{1}{2} \begin{pmatrix}
            R^0\sum_{i=0}^{N-1}||x_i + y_i||^2\\
            R^1\sum_{i=0}^{N-1}||j(x_i -y_i)||^2
        \end{pmatrix} \\
        &\propto\frac{1}{2} \begin{pmatrix}
            \sum_{i=0}^{N-1}R^0(x_i + y_i)^2\\
            \sum_{i=0}^{N-1}R^1(x_i -y_i)^2
        \end{pmatrix}.
    \end{aligned}
\end{equation}
$R^0$ and $R^1$ are the responsivities of right and left PD.
Balanced photodetection ($R^0=R^1=R$) is employed for
subtraction between $I^{0}$ and $I^{1}$, producing the final output current as,
\begin{equation}
    \label{eq:power_pair_sub}
    \begin{aligned}
    I_{o} &\propto R^0\sum_{i=0}^{N-1}(x_i + y_i)^2 - R^1\sum_{i=0}^{N-1}(x_i - y_i)^2 \\
    &\propto R (\sum_{i=0}^{N-1}(x_i + y_i)^2 - \sum_{i=0}^{N-1}(x_i - y_i)^2 )\propto R (\sum_{i=0}^{N-1}x_iy_i)
    \propto \mathbf{x} \cdot \mathbf{y}.
    \end{aligned}
\end{equation}
The differential photocurrent naturally cancels out the quadratic terms and carries the dot-product of $\mathbf{x}$ and $\mathbf{y}$.
Note that $\mathbf{x}$ and $\mathbf{y}$ for \oDot can be arbitrary vectors without any restriction.

\input{fig_tex/photonic_arch}

Our \oDot leverages coherent
light interference and WDM technique to enable general full-range vector dot-product.
Unlike prior PTC designs~\cite{NP_NATURE2017_Shen, NP_SciRep2017_Tait}, our \oDot can \emph{simultaneously} support the following three critical features:
\ding{202} \textbf{Support full-range inputs and outputs}. 
Given the coherent interference mechanism of \oDot,
the signs of inputs can be encoded to phases using MZM to achieve full-range encoding.
As shown in Section~\ref{subsec:basics}, the electric field at the output port of MZM $E_{out}=E_{in}\cos\phi$
that allows for a full encoding range of $[-1, 1]$ by tuning $\theta \in [0, \pi]$. 
Besides, balanced photodetection enables the detection of full-range outputs as positive or negative photocurrent.
Different from incoherent designs~\cite{NP_SciRep2017_Tait, NP_DAC2021_Sunny, NP_ISCA2021_shiflett} that require decomposing full-range operands into two non-negative ones and processing separately, we inherently support full-range operands at one shot with no extra overhead.
\ding{203} \textbf{Support for dynamic operand}. Both operands are optically encoded at high speed ($\sim$10 ps), thus, can be flexibly switched/reprogrammed without causing any latency bottleneck, unlike previous weight-static PTCs~\cite{NP_NATURE2017_Shen, NP_Nature2021_Feldmann}, a crucial feature to support attention.
\ding{204} \textbf{Superior computing parallelism and fully-passive computing core}. \oDot leverages WDM to enable spectral parallelism such that different wavelengths can share the same \oDot unit, achieving superior computing density and parallelism. 
The PS and DC in \oDot are entirely passive/fixed, resulting in zero energy consumption, no external control overhead, and no thermal crosstalk concern.

\subsection{DPTC: Dynamically-Operated Photonic Tensor Core}
To support MM using optical dot-product engines, 
some accelerators~\cite{NP_DAC2021_Sunny, NP_arxiv2023_Afifi, NP_NatureCom2022_Mourgias} directly map each dot-product of MM onto the dot-product engine.
However, it incurs nontrivial signal modulation overhead as operand sharing is limited.

Therefore, we present a novel photonic tensor core design, named \oMM, by constructing a compact crossbar-style array of \oDot units to \textbf{maximize the intra-core operand sharing} to largely reduce operand modulation cost, shown in Figure \ref{fig:photonic_arch}(b). 
This design enables the efficient sharing of photonic waveguide buses across \oDot units and facilitates ultra-parallel MM.
A $N_v \times N_h$ \oMM consists of $N_v \times N_h$ \oDot units, where $N_v$ and $N_h$ represent the numbers of input waveguide along the vertical and horizontal directions, respectively.

\noindent\textbf{WDM signal modulation unit(Figure~\ref{fig:photonic_arch}(b) \textcolor{pink}{pink} region)}:
The optical inputs are driven by coherent sources with phase shifters to control phases.
Each waveguide in \oMM has a WDM signal modulation unit where total $N_{\lambda}$ wavelengths are separated by a WDM DEMUX, individually modulated by high-speed MZMs, and merged in one waveguide by a WDM MUX.

\noindent\textbf{Intra-core optical broadcast unit (Figure~\ref{fig:photonic_arch}(b) \textcolor{citecolor}{green} region)}:
To largely reduce the signal modulation cost, each modulated WDM signal is broadcast to a row or a column of \oDot units through the intra-core optical broadcast unit.
A copy of the input vectors is coupled out from the optical bus and fed into the \oDot, enabling operand sharing and thus largely reducing modulation overhead in our dynamically-operated design. 
Specifically, for a $[N_h, N_{\lambda}] \times [N_{\lambda}, N_v]$ MM workload, the DAC and MZM modulation cost of our \oMM is
\begin{equation}
    E_{\text{encode}} \approx (N_hN_{\lambda} + N_{\lambda}N_v)(E_{\text{DAC}} + E_{\text{MZM}}).
\end{equation}
Compared to prior work that simply utilizes separate vector dot-product engines to implement MM without enabling operand sharing~\cite{NP_DAC2021_Sunny, NP_arxiv2023_Afifi, NP_NatureCom2022_Mourgias}, whose encoding cost is $(2N_hN_vN_\lambda)(E_{\text{DAC}} + E_{\text{MZM}})$, 
the intra-core optical broadcast helps \oMM save $(2N_hN_v)/(N_h + N_v)\times$ encoding cost.
For instance, when $N_h=N_v=N_{\lambda} = 12$, \oMM shows 12$\times$ less encoding cost.
This is one key reason why our design can still preserve high energy efficiency even if we need dynamically modulate both operands.

To summarize, \oMM inherits the ability to support full-range dynamic operands from \oDot and incorporates a more compact and energy-efficient crossbar-style design.
Unlike most previous designs supporting only MVM, our \oMM enables one-shot MM with ultra-high computation parallelism.
Moreover, leveraging the broadcast ability of light, we maximize the intra-core sharing of modulated signals among multiple \oDot units to largely amortize the operand encoding cost.

\subsection{Robustness Analysis of Proposed Photonic Design}
\label{subsec:robustness}
Analog optical computing systems are subject to various noises, e.g., encoding noise, WDM dispersion, and device imperfection.
Here, we analyze the noise impact on fundamental \oDot unit and show its inherent robustness to variations.

\noindent\textbf{Optical encoding noise}.~
In \oDot, both operands are encoded as optical signals, which are inevitably susceptible to encoding noise, i.e., stochastic magnitude and phase drift.
Specifically,
consider two optical operands $x$ and $y$, we have $x' = \hat{x}_i e^{j\delta\phi_{x}} = (x + \delta x)e^{j\delta\phi_{x}}$ and $y' = \hat{y}_ie^{j\delta\phi_{y}} = (y + \delta y) e^{j\delta\phi_{y}}$, where $\delta x$ and $\delta y$ denote the magnitude drift, and $\delta\phi_{x}$ and $\delta\phi_{y}$ denote the phase drift.
For simplicity, we extract the relative phase drift between the two operands and express it as an equivalent phase drift $\delta\phi_d = \delta\phi_y -\delta\phi_x$, 
following a Gaussian distribution $\delta\phi_d \sim \mathcal{N}(0, \sigma_\phi ^2)$.
The magnitude drift follows a Gaussian distribution $\delta x\sim\mathcal{N}(0, (\sigma_x|x|)^2)$, where the standard deviation depends on the absolute value $|x|$ we want to encode.
With the encoding noise, the noisy transfer function of \oDot is expressed as,
\begin{equation}
    \label{eq:signal_pair_noise}
    \begin{aligned}
       \begin{pmatrix}
            z_i^0\\
            z_i^1
        \end{pmatrix}
       &=\frac{1}{\sqrt{2}} \begin{pmatrix}
            1 & j\\
            j & 1
        \end{pmatrix} 
        \begin{pmatrix}
            1 & 0\\
            0 & e^{-j\pi/2}
        \end{pmatrix}
        \begin{pmatrix}
            \hat{x}_i \\
            \hat{y}_i e^{j\delta\phi_{d_i}}
        \end{pmatrix} \\
        &=\frac{1}{\sqrt{2}}\begin{pmatrix}
            (\hat{x}_i - \sin\phi_i \hat{y}_i) + j\hat{y}_i\cos\phi_i \\
            \hat{y}_i\cos\phi_i + j(\hat{x}_i +\sin\phi_i\hat{y})        
            \end{pmatrix},
    \end{aligned}
\end{equation}
where $\phi_i = \delta\phi_{d_i} - \pi / 2$ as a perturbed value around 
$-\pi/2$.

\input{fig_tex/sweep_lambda}

\noindent\textbf{WDM dispersion.}~Our architecture leverages WDM to allow multiple wavelengths to share the same \oDot unit.
Nevertheless, even with the adoption of broadband devices (coupler, phase shifter),
photonic circuits still exhibit slightly different responses to different wavelengths. 
Specifically, the wavelength-dependent transfer function of the directional coupler is described by {\scriptsize$\begin{pmatrix}
            t & kj\\
            kj & t
        \end{pmatrix}$},
where $t = \sqrt{1-\kappa(\lambda)}$ and $k = \sqrt{\kappa(\lambda)}$.
$\kappa(\lambda)$ is the wavelength-dependent power coupling factor and computed as $\kappa(\lambda) = \sin^2((\pi L_{c}(\lambda_0))/ 4L_{c}(\lambda))$, where $L_c(\lambda)$ is the  $100\%$ coupling length.
$\kappa(\lambda_0)$ is designed to be $1/2$.
Besides, the phase response of the phase shifter $\Delta \phi(\lambda) = 2\pi \Delta n_{\text{eff}}L/\lambda$, which is also wavelength-dependent.
In this paper, we follow Dense WDM standard \cite{NP_DWDM} with a 0.4 nm wavelength channel spacing and choose the center wavelength $\lambda_0=1.55$ nm.
We sweep 25 wavelengths around $\lambda_0$ and show the corresponding $\kappa$ and $\phi$ in Figure~\ref{fig:sweep_lambda}.
The maximum relative difference of $\kappa$ between $\lambda_0$ and the furthest wavelength is $\sim$$1.8\%$.
The maximum dispersion-induced phase difference is $0.28^{\circ}$, which is negligible compared to the $360^{\circ}$ MZM tuning range.

By considering the impact of WDM dispersion, we have the wavelength-dependent transfer function of \oDot as
\begin{equation}
  \label{eq:signal_pair_noise_wdm}
    \begin{aligned}
       \begin{pmatrix}
            z_i^0\\
            z_i^1
        \end{pmatrix}
        &=\begin{pmatrix}
            (t_i\hat{x}_i - k_i\hat{y}_i\sin\phi_i) + jk_i\hat{y}_i\cos\phi_i \\
            t_i\hat{y}_i\cos\phi_i + j(k_i\hat{x}_i +t_i\hat{y}\sin\phi_i)        
            \end{pmatrix}
    \end{aligned},
\end{equation}
where $\phi_i = \delta\phi_{d_i} - \pi / 2 + \delta\phi_{\lambda_i}$ including dispersion induced phase drift $\delta\phi_{\lambda_i}$ in radian mode.
Hence, the output photocurrent of \oDot, considering both encoding noise and dispersion, is
\begin{equation}
\label{eq:power_pair_sub_noise_wdm}
    \begin{aligned}
    I_{o}  &\propto \sum_{i=0}^{N-1}( \frac{(k_i^2 - t_i^2) \hat{x}^2_i + (t_i^2 - k_i^2) \hat{y}^2_i}{2} + 2t_ik_i\hat{x}_i \hat{y}_i (-\sin\phi_i)) \\
    &\propto \sum_{i=0}^{N-1} (2 \underbrace{k_i\sqrt{1-k_i^2}(-\sin\phi_i)}_{\text{\ding{202}multiplicative noise }}\hat{x}_i \hat{y}_i + \underbrace{(2k_i^2 - 1)\frac{ \hat{x}^2_i - \hat{y}^2_i}{2}}_{\text{\ding{203}additive noise}}) \\
    \end{aligned},
\end{equation}
where each scalar product $x_iy_i$ has non-ideal scaling factors $\sin\phi_i$ and $2k_i \sqrt{1-k_i^2}$, and an additive noise relative to $\hat{x}_i^2 - \hat{y}_i^2$.

\input{fig_tex/system_arch}

\noindent\textbf{\oDot is inherently robust to \ding{202}~multiplicative noise}, as our design points $k = \frac{1}{\sqrt{2}}$ and $\phi=-\frac{\pi}{2}$ are at the \emph{local optima} of $x\sqrt{1-x^2}$ and $\sin(\cdot)$ with minimized sensitivity to perturbations.

\noindent\textbf{\oDot is robust to \ding{203}~additive noise}.~
Our photonic design is also robust to the additive error as it can be naturally \emph{canceled out}.
In our \oDot engine, MZM modulation range is $[-1, 1]$. 
Hence, before mapping $\mathbf{x}$ and $\mathbf{y}$ onto \oDot, we need to scale them into [-1, 1] with their maximum absolute value $\beta_x = \max(|x|)$ and $\beta_y = \max(|y|)$.
This normalization effect naturally ensures $\hat{x}_i$ and $\hat{y}_i$ in Eq.~\eqref{eq:power_pair_sub_noise_wdm} in a similar range, i,e., [-1, 1], such that in a vector dot-product, the introduced multiple $\hat{x}_i^2$ and $\hat{y}_i^2$ can be self-compensated.
Moreover, the square effect and scaling factor $\frac{1}{2}$ largely reduce the additive noise. 

Overall, our design shows remarkable robustness to phase drift and WDM dispersion. 
The exceptional noise tolerance to dispersion enables \oDot to scale with a large number of wavelengths, significantly enhancing the spectral parallelism.

\noindent\textbf{Other Noises}.~
To further simulate other noises in the system (\oMM), e.g., photo-detection noise, imperfect coupling ratios of direction couplers, we generally add a multiplicative noise to the computation outputs of \oMM, $\hat{I_o}=I_o(1+\epsilon)$, where $\epsilon\sim \mathcal{N}(0, 0.05^2)$, when simulating the accuracy of running Transformer on our photonic designs.

%% file: fig_tex/photonic_arch.tex
\begin{figure*}
    \centering
\includegraphics[width=0.9\textwidth]{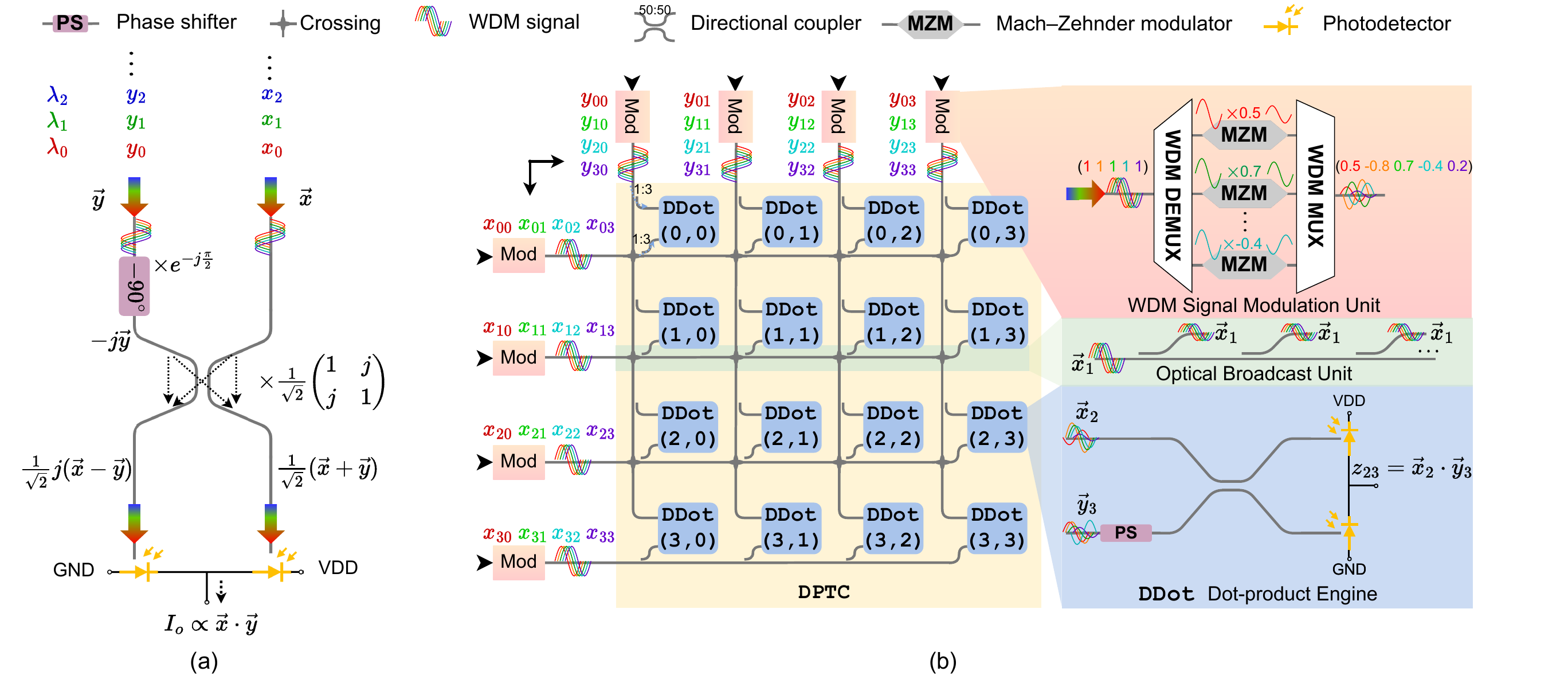}
    \caption{~
    (a) The proposed \oDot dot-product engine. Multi-wavelength signals propagate concurrently on the waveguide.
    (b) The proposed \oMM matrix-matrix multiplication unit with input WDM signals broadcasting. 
    }
    \label{fig:photonic_arch}
    \vspace{-10pt}
\end{figure*}

%% file: fig_tex/sweep_lambda.tex
\begin{figure}
    \centering
    \vspace{-5pt}
    \subfloat[]{
        \includegraphics[width=0.22\textwidth]{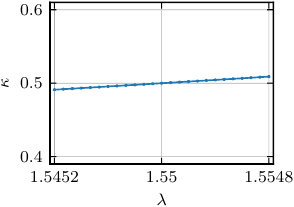}
        \label{fig:kappa_design_point}
    }
    \subfloat[]{
        \includegraphics[width=0.24\textwidth]{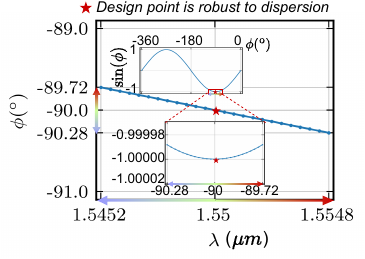}
        \label{fig:sin_phi_design_point}
    }
    \vspace{-5pt}
    \caption{~\small Our design point is robust to non-ideal dispersion effects.
    Coupling coefficient $\kappa$ and phase shift $\phi$ are not sensitive to wavelength-dependent device responses (i.e., dispersion).
    }
    \label{fig:sweep_lambda}
    \vspace{-10pt}
\end{figure}

%% file: fig_tex/system_arch.tex
\begin{figure*}
    \centering
    \includegraphics[width=0.9\textwidth]{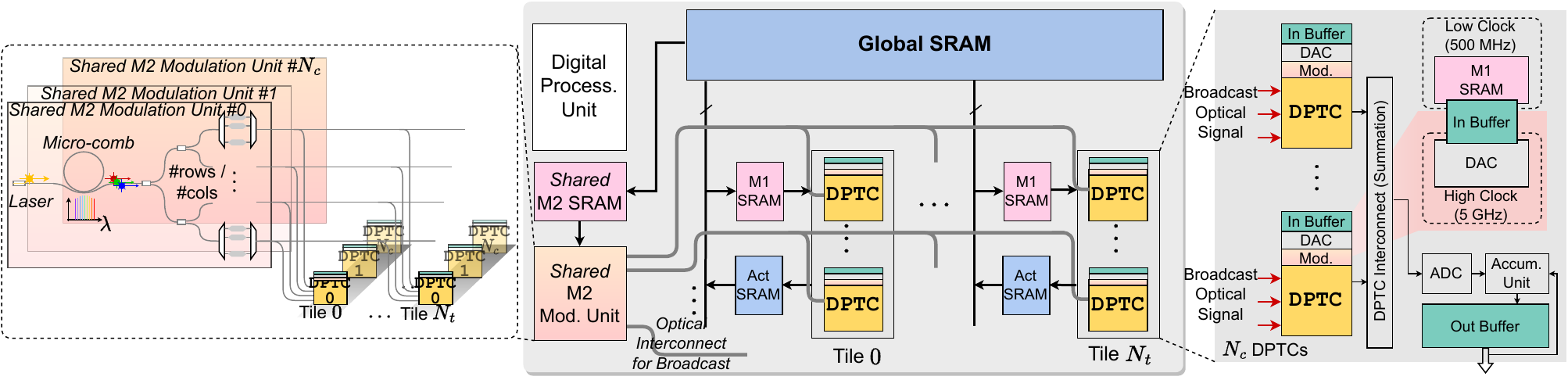}
    \vspace{-5pt}
    \caption{~
    High-level architecture of the proposed \name.
    It has a three-level memory hierarchy, multiple photonic analog computing tiles/cores, on-chip multi-wavelength light sources, and optical interconnects for data broadcast.
    }
    \label{fig:system_arch}
    \vspace{-10pt}
\end{figure*}

%% file: doc/arch.tex
\section{Proposed Photonic Accelerator Architecture}
In this section, we present the high-level architectural design of the proposed accelerator, \name.

\subsection{Overall System Design}
The \name architecture
contains analog photonic computing units for General matrix multiplication (GEMM) acceleration, photonic interconnect for data broadcast, and underlying electronics for other operations, including signal conversion, data storage, nonlinearity functions, and SoftMax.

\noindent \textbf{Architecture overview.}~
Figure \ref{fig:system_arch} illustrates the overall micro-architecture with a particular emphasis on the photonic part.
We incorporate multiple photonic tensor cores (\oMM) into a single chip to increase the amount of parallel processing.
We cluster $N_c$ \oMM into one tile, and we have $N_t$ tiles in total in the accelerator.
All the photonic tensor cores \oMM are clocked at 5GHz for a conservative assumption.
The operands of each \oMM undergo an electrical-to-optical conversion (E/O) with digital-to-analog converters (DAC), 
and the outputs of \oMM need analog-to-digital converts (ADC) to bring the analog result back to the digital domain.
We have digital processing units to process non-GEMM operations needed in Transformer.

\noindent\textbf{Memory}.
For the memory part, we have a large 2MB global SRAM, which interacts with DRAM and is responsible for holding inputs, activations, and weights.
The size of the global SRAM is designed to be sufficient for (1) storing single-layer largest activations for targeted low-bit BERT/DeiT Transformers' single-batch inference.
(2) double buffering for required off-chip data that is loaded chunk by chunk based on the tiling algorithm in Figure~\ref{fig:mm_tiling}
to overlap data transfer time with computation.
The corresponding off-chip HBM bandwidth is set to ensure the data access latency is hidden by computation latency.
In addition, each tile has its own 4KB SRAM to hold the operands for its own matrix multiplication workload, as well as an activation SRAM to store the computed results. Since the photonic part of \name runs at a clock speed of 5GHz, and our \oMM unit handles one small matrix multiplication in one clock cycle, the required data bandwidth is high. 
To address this, we follow~\cite{NP_arxiv2021_Demirkiran} and decouple the large SRAM array into smaller 32KB sub-arrays. This improves data bandwidth by reading data from these sub-arrays with a small shifting interval.
Data buffers are used for each photonic tensor core to communicate the DACs and SRAM in different clock domains.

\name uses output-stationary (OS) dataflow and proposes multiple architecture-level optimization techniques
to efficiently support GEMM operations, which will be elaborated on later.

\input{fig_tex/mm_tiling}
\noindent\textbf{Photonic tensor cores}.~
Photonic tenor core \oMM works with necessary digital components, including DACs, ADCs, TIAs, and data buffers and operates at $4$-bit precision by default.
\oMM can implement $[N_h, N_{\lambda}] \times [N_{\lambda}, N_v]$ matrix multiplication in one cycle.
Large GEMM operations in Transformer are mapped onto \oMM by using tiled matrix multiplication.

\noindent\textbf{Photonic Interconnect}.
We leverage photonic interconnect to broadcast optical signals across photonic tensor cores when operand sharing opportunity exists.

\noindent\textbf{Digital processing units}.~
We assume all other non-GEMM operations are implemented using digital electronics.

\input{tablels/params}

\subsection{Dataflow}
Our dynamically-operated \oMM frees the selection of dataflow by having both operands dynamically encoded.
Most prior photonic accelerators typically map one operand into photonic circuit states that cannot be dynamically switched due to slow reconfiguration speed.
This design concept limits its dataflow selection, making them only suitable for weight-stationary~(WS) dataflow~\cite{NP_arxiv2021_Demirkiran}.
Note that the WS stationary dataflow in photonic accelerators differs from that of the digital electronics accelerator.
They are constrained to only favor the WS dataflow, while the normal WS stationary dataflow is a design choice to explore data locality and data reuse.

To efficiently support GEMM on our multi-tile accelerator, 
we consider fine-grained tiling and carefully design the spatial/temporal mappings, as illustrated in Figure~\ref{fig:mm_tiling}.
We use the OS dataflow as the basic design principle for calculating MM between $\mathbf{M_1}$ and $\mathbf{M_2}$.
The OS dataflow enables us to minimize the on-chip buffer size \cite{HWA_HPCA2023_You, HWA_MICRO2021_Lu} and adopt customized optimization techniques to reduce the cross-domain signal conversion overhead in our mixed-signal accelerator.
Specifically, we tile matrix $\mathbf{M_1}$ along the $N_1$ dimension and map them to different tiles spatially.
Each tile is responsible for calculating the multiplication result between one tiled row of $\mathbf{M_1}$ and matrix $\mathbf{M_2}$ temporally.
At each cycle, each tile handles tiled matrix multiplication.
Since we have multiple \oMM in each spatial unit, we distribute the tiled MM workload among these cores. 
The outputs of multiple cores are first accumulated by photocurrent summation in the analog domain, followed by A/D conversion. 
Then, the partial sums are sent to the output buffer for further sequential accumulation in the digital domain.
This analog domain partial summation can not only reduce the A/D conversion cost but also avoid the precision loss during A/D conversion with full-precision photocurrent summation.

\subsection{Architecture-Level Optimization}
\label{subsec:opt}
The expensive electrical-to-optical (E-O) conversion and optical-to-electrical (O-E) costs remain to be the key bottleneck for emerging photonic systems, which is also true in \name.
Therefore, we consider the two following optimization opportunities to reduce the E-O and O-E costs.

\subsubsection{Inter-core Operand Broadcast via Optical Interconnect}
As shown in Figure~\ref{fig:mm_tiling},
different tiles process the multiplication between different chunks of $\mathbf{M}_1$ and the same chunk of $\mathbf{M}_2$.
This creates an opportunity to share the common $\mathbf{M}_2$ part across multiple tiles. 
Leveraging the exceptional signal broadcast capabilities of optics, we encode the shared $\mathbf{M}_2$ in optical signals using global modulation units (as shown in the left part of Figure~\ref{fig:system_arch}). 
These modulated signals are efficiently broadcasted to different \oMM units via optical interconnects, leading to an architecture-level $N_t\times$ reduction in data movement and modulation costs.

\subsubsection{Analog-Domain Temporal Accumulation with Time Integral}
As our \oMM design supports efficient dynamic operand switching, we adopt OS dataflow that can enable analog-domain temporal accumulation~\cite{NP_arxiv2023_Zheng, NP_HPCA2023_Li} via time integral to reduce the ADC overhead.
This technique is not applicable to prior WS-static accelerators since their outputs are accumulated to different buffer locations across time steps, preventing utilizing temporal accumulation.
Analog temporal accumulation accumulates results by using photodetectors and capacitors to store charges, which can be read at a later time. 
This allows ADC to operate at a lower frequency, thereby reducing ADC costs.
Additionally, by performing accumulation before signal digitization, the partial summation is achieved in the analog domain in full precision, leading to accuracy~\cite{NP_HPCA2023_Li}.
In this paper, we set temporal accumulation depth to 3, i.e., A/D conversion happens once every three analog accumulation steps.

%% file: fig_tex/mm_tiling.tex
\begin{figure}
    \centering
\includegraphics[width=0.9\columnwidth]{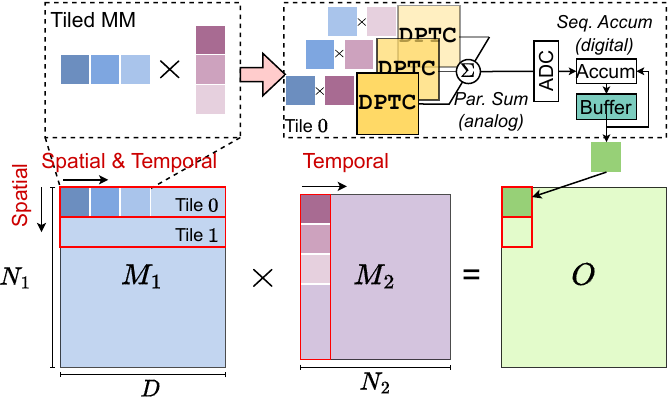}
    \caption{~
    Tiling and spatial/temporal mapping for processing GEMM. $M_1$ is the weight matrix when processing the linear layer, which is loaded chunk by chunk off-chip.
    }
    \label{fig:mm_tiling}
\end{figure}

%% file: tablels/params.tex
\begin{table}[ht]
\centering
\caption{Table of notations used in the accelerator design}
\label{tab:arch_params}
\resizebox{0.95\columnwidth}{!}{
\begin{tabular}{c|c}
\hline
Notation      & Definition                                                 \\ \hline
$N_h$         & \# input horizontal waveguides of each \oMM \\
$N_v$         & \# input vertical waveguides of each \oMM   \\
$N_{\lambda}$ & \# wavelengths of each \oMM                 \\
$N_{c}$       & \# photonic tensor cores in each tile                     \\
$N_{t}$       & \# tiles in our accelerator                         \\ \hline
\end{tabular}
}
\vspace{-10pt}
\end{table}

%% file: doc/6result.tex
\input{tablels/photonic_device_area}

\section{Evaluation}
\label{sec:ExperimentalResults}

\subsection{Evaluation Setup}
\noindent\textbf{System setup}.~
We build a Python-based simulator to simulate the latency, power, area, and energy efficiency of our proposed accelerator on actual Transformer inference.
The area, leakage power, and access energy of the memory system are modeled using PCACTI~\cite{HWA_AVLSI2014_Shafaei} in 14 nm.
We use high-bandwidth memory (HBM) to supply data to the photonic system with a bandwidth $>$1TB/s~\cite{DRAM}.
The energy cost of digital processing units is obtained from~\cite{HWA_APCCAS2018_Wang, HWA_ICCD2022_Prashanth, guo2020att}.
We use the ADC~\cite{NP_ISSCC2022_Liu} and DAC~\cite{NP_VLSI2020_Caragiulo} with similar technology nodes (14 nm and 16 nm), while their bit-widths and frequency do not precisely match our settings.
We follow~\cite{HWA_ISLPED2018_Kim} to scale the ADC and DAC power according to the bit-width and frequency requirement of the photonic computing units.
Table~\ref{tab:device} lists the parameters of used photonic devices.
The laser power is set to meet the minimum power requirement of the photodetector considering system loss~\cite{NP_DAC2021_Sunny} and is scaled based on the precision requirement and wall-plug efficiency.
During simulation, the simulator implements the detail of the tiling algorithm and uses a batch size of 1.
The deep pipeline of the photonic/digital processing unit is not adopted in this paper, which can be employed to further improve the system performance~\cite{NP_arxiv2021_Demirkiran}.

\input{fig_tex/lumerical}

\noindent\textbf{Functionality validation}.
We use Lumerical INTERCONNECT tools~\cite{NP_lumerical} with the AIM process design kit (PDK) to validate the functionality of our \oDot engine.
We inject optical encoding noise as discussed in Section~\ref{subsec:robustness} with input noise std. and phase noise std. being 0.03 and 2$^\circ$, respectively.
WDM dispersion is naturally considered in the simulation.
12 wavelengths are used with 0.4 nm wavelength channel spacing.
The dot-product error of one random length-12 dot-product is 2.6\% and 3.4\% in 4-bit and 8-bit precision, respectively, 
as shown in Figure~\ref{fig:lumerical}.

\noindent\textbf{Models, datasets, training, and inference settings}.
We use DeiT~\cite{NN_ICML2021_Touvron} and BERT~\cite{NN_arxiv2018_Devlin} to evaluate the efficiency and accuracy.
Both are well-recognized Vision and NLP Transformers.
We apply low-bit quantization on both weight and activation~\cite{NN_ICLR2020_Esser}.
Noise-aware training is applied with encoding and systematical noise injected.
We model the computation by using noisy analytic transformation (Eq.~\eqref{eq:power_pair_sub_noise_wdm}) with
all noises in Section~\ref{subsec:robustness} injected to
test the inference accuracy.

\input{tablels/dota_params}

\noindent\textbf{Architecture configurations}.
We design two versions of \name called \sname-B (base version) and \sname-L (large version) with detailed parameters in Table~\ref{tab:dota_config}.
\sname-B has four tiles, where each tile has two \oMM.
\sname-L doubles the number of tiles to increase its throughput for large models.

\input{fig_tex/system_area}
\input{fig_tex/system_power}
\input{fig_tex/fig_ptc_scaling}

\subsection{System Efficiency Analysis}

\noindent\textbf{Area breakdown}.~ Figure~\ref{fig:DotaArea} shows the area breakdown of our \sname-B and \sname-L architectures. The \sname-B has 60.3 $\text{mm}^2$, which is around half of the area of the \sname-L architecture (112.82 $\text{mm}^2$). For each architecture, the photonic core, memory, and DAC contribute the largest portion of the area, with around 20\%, 25\%, and 25\% share, respectively. The remaining components, such as laser, ADC, and MZM, account for less than 30\% of the overall area.

\noindent\textbf{Power breakdown}.~Figure~\ref{fig:system_power} shows the power breakdown for \sname-B for two data precision settings, 4-bit and 8-bit.
The \sname-L shares a similar power breakdown with 28.06 W (4-bit) and 95.92 W (8-bit) in total.
In our \name, all operands are dynamically encoded, resulting in a dominant operand encoding cost (DAC and MZM).
The reported operand modulation cost has been optimized through operand sharing at intra-core and inter-core levels. 
The 8-bit \sname-B consumes more than three times the power of the 4-bit one. 
This increase is primarily attributed to the largely increased power of high-bit DACs, which account for over 50\% of the overall power in the 8-bit architecture. 
Also the laser power also significantly increases from 0.77 W to 12.3 W to satisfy higher output precision.

\noindent\textbf{Area, power, latency, and performance scaling}.~ In Figure~\ref{fig:ptc_scaling}, we show how the area, power consumption, and latency scale with the size of \oMM. 
Here, we don't consider input to be globally modulated, i.e., DACs are not shared across different tiles for better observing the scaling effect.
The area increases from 5.9 $mm^2$ to 49.3 $mm^2$ when increasing the core size from 8 to 32. 
The ratio of each part roughly remains the same. 
Power consumption of one single core increases from 1.1 W to 17 W as the core size increases from 8 to 32, and the modulation, ADC, and DAC take the lion's share of the overall power consumption. 
The latency scaling shows a different pattern from the area and power consumption. 
The optics latency increases approximately linearly with the size as the optical path increases.
The EO/OE latency remains almost the same. 
As shown in Figure~\ref{fig:ptc_perf_scale}, we further show the performance and efficiency scaling of our design. 
As the core size increases, the performance (TOPS), area efficiency (TOPS/mm$^2$), and energy efficiency (TOPS/W) increase while the energy efficiency per unit area (TOPS/W/mm$^2$) decreases due to the bottleneck of ADCs and DACs.

\noindent\textbf{Wavelength scaling}.~
Thanks to the superior robustness of our photonic design to WDM dispersion effects,
we can increase the number of wavelengths to further boost processing parallelism.
In our system, we use Microdisk~\cite{NP_NatureComm2014_Timurdogan} as the filter to implement the WDM MUX and WDM DEMUX.
However, the microdisk imposes a free spectral range (FSR), which limits the number of wavelengths.
Given its FSR=5.6 THz
and design wavelength at 1550 nm,
we can have its wavelength range as
\begin{equation}
    \begin{aligned}
        \lambda_{l} = c/(f_0+FSR/2) = 1527.88 \ \text{nm} ;\\
        \lambda_{r} = c/(f_0-FSR/2) = 1572.76 \ \text{nm}.
    \end{aligned}
\end{equation}
With a 0.4 nm channel spacing, we have up to 112 wavelengths.

\subsection{Compare to State-of-the-art Photonic Accelerators}

\underline{Baseline}:
We have built baseline systems based on incoherent MRR bank~\cite{NP_SciRep2017_Tait} and coherent MZI array~\cite{NP_NATURE2017_Shen}. 
For a fair comparison, we scale the number of PTC in baselines to
match area of our base version of \name (\sname-B).
Since the MZI array cannot efficiently support dynamic MM, we assume to use MRR bank to implement the MHA part for it.
We adopt weight-static dataflow for baselines.

\underline{Energy cost model of PTC execution}:
Consider a GEMM between two matrices, $\mathbf{X}\in \mathbb{R}^{m\times d}$ and $\mathbf{Y}\in\mathbb{R}^{d\times n}$, and a PTC performing $[k, k]\times [k, k^{'}]$ MM at one time where $k^{'} =1$ when PTC supports MVM.
$k$ is typically small compared to the matrix dimension. 
Thus, tiled matrix multiplication is needed for large GEMM operations, executing PTC $T=\lceil \frac{m}{k} \rceil \lceil \frac{d}{k} \rceil \lceil \frac{n}{k^{'}} \rceil$ times.
If the energy to load (detect) one scalar input (output) is $E_{\text{load}}$ ($E_{\text{det}}$) and the laser energy per cycle is $E_{\text{laser}}$, the PTC computation energy cost~\cite{NP_ICCAD2021_Li} (w/o data-movement cost) is:
\begin{equation}
\label{eq:energy}
    \begin{aligned}
        E&= T\cdot E_{\text{laser}} + \underbrace{T\cdot k^2 E_{\text{load}}}_{\text{load X}} + \underbrace{T\cdot kk^{'} E_{\text{load}}}_{\text{load Y}} + \underbrace{T\cdot kk^{'} E_{\text{det}}}_{\text{output}} \\
        &\approx T\cdot E_{\text{laser}} +md\lceil \frac{n}{k^{'}} \rceil(E_{\text{DAC}} + E_{\text{mod}}) \\
        +&dn\lceil \frac{m}{k} \rceil(E_{\text{DAC}} + E_{\text{mod}})+ \lceil \frac{d}{k} \rceil mn (E_{\text{PD}} + E_{\text{amp}} + E_{\text{ADC}}).
    \end{aligned}
\end{equation}
$E_{\text{load}}$ contains the energy costs of DAC ($E_{\text{DAC}}$) and signal modulation ($E_{\text{mod}}$).
$E_{\text{det}}$ represents the cost of detecting optical signal ($E_{\text{PD}}$), amplifying it ($E_{\text{amp}}$), and performing analog-to-digital conversion ($E_{\text{ADC}}$).
Note that in weight-static designs (MRR bank and MZI array),
the DAC and dynamic modulation cost of the static operand can be amortized.
However, MRR bank incurs an additional mW-level static locking power~\cite{NP_arxiv2021_Demirkiran} to maintain the encoded value in the resonator, which cannot be amortized. 
The total locking cost scales with the total number of computations, i,e., $mdn$.
\input{fig_tex/fig_ptc_perf_scale}

\input{fig_tex/fig_attention_linear}

\underline{Comparison on an Attention}:
In Figure~\ref{fig:compare_linear_attention}(left), we compare our \sname-B and MRR baseline on an example MHA workload ($\mathbf{Q} \mathbf{K}^T$) in DeiT-T.
We disable the architecture-level optimization of \sname-B (i.e., photocurrent summation within the same tile, inter-core operand broadcast, and analog domain temporal accumulation) to purely compare the PTC design.
We denote this version of \sname-B as \sname-\texttt{crossbar}-B.
Although MRR bank can amortize the DAC cost of the static op1
the unamortized static operand locking power (\textcolor{mod1color}{op1-mod}) contributes to $>$$40\%$ of total energy cost.
Moreover, its inability to support full-range inputs requires decomposing them into non-negative ones and processing them separately(double $T$ in Eq.~\eqref{eq:energy}).
The induced  2$\times$ input encoding (\textcolor{DAC2color}{op2-DAC}, \textcolor{mod2color}{op2-mod}) and detection (\textcolor{detcolor}{det}, \textcolor{ADCcolor}{ADC}) cost eliminates the weight-static benefit.
Our \sname-\texttt{crossbar}-B has a 2.62$\times$ less energy cost with inherent full-range input support, no locking power, and maximized intra-core operand sharing to amortize encoding costs for both operands.

\input{tablels/compare_prior_onn}

\underline{Comparison on a Linear layer}:
We highlight that our \name is more efficient than prior designs, even on weight-static linear layer workload.
In Figure~\ref{fig:compare_linear_attention}(right), we compare the energy cost breakdown between MRR, MZI baselines, and \sname-\texttt{crossbar}-B (\sname-B without architecture-level optimization) on the first linear layer of FFN block in DeiT-T.
We emphasize that our is not merely designed for attention but for generic GEMM acceleration supporting ultra-parallel and energy-efficient full-range MM.
Even though our design has two operands ($W$ and $x$) being dynamically encoded, the resultant extra encoding cost (\textcolor{mod1color}{op1-mod} and \textcolor{DAC1color}{op1-DAC}) for weight operand is marginal in our total energy since of our topology-enabled intra-core sharing. 
Hence, the efficiency superiority of \name still holds and is well justified.
The main reason that leads to the significantly worse energy efficiency of MZI-based PTC is its prohibitive laser power (\textcolor{lasercolor}{laser}) due to the huge insertion loss of deeply cascaded MZI array, which takes over 75\% of its total energy cost.
Note that its laser energy itself is already 2.9$\times$ higher than our total energy.
Moreover, given the limited bandwidth and bulky footprint of the MZI array, we can only fit a few cores on the chip, each supporting MVM using a single wavelength.
Thus, it consumes much more cycles (longer latency) to finish an MM workload than our ultra-parallel, compact \oMM design.
Based on the above two reasons, even if the weight-static MZI array sounds more efficient on the linear layer workloads than our dynamic tensor core, our \name shows surprisingly better energy efficiency than it even on linear layer workloads.
This conclusion seems counterintuitive but well-explained here and evidenced by the quantitative analysis in Figure~\ref{fig:compare_linear_attention}(right).

\input{fig_tex/fig_detail_breakdown_compare_mrr}

\underline{Comparison of \name variants}:
Figure~\ref{fig:compare_linear_attentio_breakdown} shows the energy cost breakdown on Attention/linear layers between MRR bank and different variants of the base \name variants (\sname-\texttt{broadcast}-B, \sname-\texttt{crossbar}-B, \sname-B) to highlight the contributions of introduced features to energy efficiency.
\sname-B activates all features in our accelerator design, i.e., crossbar-topology and architecture-level optimization.
\sname-\texttt{crossbar}-B turns off the architecture-level optimization while preserving our crossbar-style topology for our photonic tensor core \oMM.  
\sname-\texttt{broadcast}-B
adopts a  similar \oMM topology to MRR, only broadcasting input operand to different dot-product units \oDot. It doesn't employ the crossbar topology to share both operands intra-core but still features dynamic, full-range operand support.

\input{fig_tex/fig_compare_energy_fps}

As in Figure~\ref{fig:compare_linear_attentio_breakdown}, MRR bank exhibits significant static operand locking power(\textcolor{mod1color}{op1-mod})  ($>$40\% of the total energy), as weight-static dataflow can only amortize dynamic power.
Although vanilla \sname-\texttt{broadcast}-B features a slightly higher energy cost than MRR due to the unshared modulation cost of op1 (\textcolor{mod1color}{op1-mod} and \textcolor{DAC1color}{op1-DAC}), the full-range feature results in $>$2$\times$ smaller energy cost for other parts(\textcolor{DAC2color}{op2-DAC}, \textcolor{detcolor}{det}, \textcolor{ADCcolor}{ADC}), as avoiding the need of running PTC twice for full-range inputs.
By adopting the crossbar topology, \sname-\texttt{crossbar}-B not only eliminates modulation overhead but also achieves over 2x better energy efficiency than MRR bank.

\sname-B achieves the lowest energy cost.
Compared to \sname-\texttt{crossbar}-B,
it further equips with inter-core operand broadcast with $\sim4\times$ less input operand modulation cost (\textcolor{DAC2color}{op2-DAC}, \textcolor{mod2color}{op2-mod}).
It also leverages signal-domain-wise locality by applying photocurrent summation across PTCs in the same tile and achieving temporal partial summation before sending data to the ADC by time-integral, thus reducing ADC energy costs (\textcolor{ADCcolor}{ADC}) cost by $\sim6\times$. 
This time-integral technique is uniquely supported in our dynamically-operated \oMM design, as it can be only applied for accelerators supporting \textit{output-stationary dataflow}.

\underline{Comparison on DeiT:}
Table~\ref{tab:compare_prior_onns} compares the base version of \name (\sname-B) with baselines on DeiT-T/B in 4-bit/8-bit precision. 
The results prove the superior performance of \name in terms of energy, latency, and energy-delay product (EDP).
Specifically, it outperforms the MZI array by over 8$\times$, 675$\times$, and 5000$\times$ in terms of energy, latency, and EDP, respectively.
Especially when scaling to high precision, the MZI baseline costs much more energy due to the exponentially increasing laser power.
When compared to MRR bank, \sname-B achieves energy, latency, and EDP savings of 2.6$\times$, 12.8$\times$, and 34.2$\times$ on the 8-bit setting.
On the 4-bit setting,  \sname-B achieves energy, latency, and EDP savings of 4$\times$, 12.8$\times$, and 51.7$\times$ over the MRR bank.
Even without architecture-level optimization in Sec.~\ref{subsec:opt}, \sname-B still saves over 2$\times$ energy compared to baselines, showing the efficiency of our \oMM design.
Besides the architecture-level optimization, 
this superiority of \name can be largely attributed to 
\ding{202} \textbf{High parallelism}. Our photonic tensor core performs ultra-parallel one-shot MM instead MVM, featuring significantly higher throughput and lower energy cost.
\ding{203} \textbf{High efficiency}.
\oMM features zero locking cost and low laser power (low insertion loss) and exploits intra-core and inter-core operand sharing to reduce modulation overhead.
\ding{204} \textbf{Natural full-range support} with no extra overhead.
\ding{205} \textbf{Dynamic operand switch} at high speed($<$10 ps). The use of low-loss phase shifters in
MZI array imposes a high latency term for \emph{reconfiguring PTC} (2 $\mu$s) to the total latency when switching weight operand for tiled MM.

\noindent\textbf{PTC-Level Takeaways}

\noindent\ding{202}~\emph{Exploring both spectral and spatial parallelism of optics brings significant performance}.
Besides spatial parallelism using multiple compact cores, a coherent broadband structure can further leverage multi-wavelength processing to fully unleash the spectral parallelism of optics, significantly increasing the computing parallelism and density.

\noindent\ding{203}~\emph{Dynamic full-range operand encoding enables versatility}. In contrast to prior weight-static designs with restricted applications, e.g., CNN with ReLU, 
a generic, highly-reprogrammable optical GEMM primitive, like our \oMM, can offer much wider applicability and higher efficiency to adapt to advanced and ever-evolving ML acceleration tasks.

\noindent\ding{204}~\emph{Leveraging the natural broadcast capability of optics maximizes hardware sharing and energy efficiency}.
The crossbar topology largely amortizes the footprint cost of photonic devices via extensive hardware sharing while inducing minimum overhead due to the superior broadcasting and interconnecting capability of photonic waveguides.

\noindent\textbf{Architecture-Level Takeaways}

\noindent\ding{202}~\emph{Combining photonic computing with photonic interconnect unleashes the power of optics}.
The integration of optical computing cores and cross-core optical interconnect allows for speed-of-light computing and communication with further operand sharing at the architecture level, simultaneously reducing the data movement and signal modulation cost.

\noindent\ding{203}~\emph{Offloading more computing to the analog domain relaxes the A/D conversion bottleneck}.
To reduce cross-domain signal conversion costs, it is critical to leverage the concept of \emph{signal-domain-wise locality}. By offloading more computations within the analog domain, such as analog partial summation and temporal accumulation employed in \name, the need for frequent A/D conversion is minimized, significantly alleviating power and latency bottlenecks from ADCs.

\input{fig_tex/fig_dispersion_accuracy}
\input{fig_tex/fig_noise_accuracy}

\subsection{Compare to State-of-the-art Digital Accelerators}
In Figure~\ref{fig:compare_energy_fps}, we compare \name to different hardware platforms to demonstrate our orders-of-magnitude performance improvements. 
Specifically, we make comparisons with (1) single Nvidia A100 GPU, (2) Intel Core i7-9750H CPU, (3) Cora Edge TPU~\cite{HWA_ISCA2023_Reidy}, and (4) FPGA accelerators AutoViT~\cite{auto-vit}, HEATViT~\cite{HWA_HPCA2023_Dong}.
For GPU and CPU, we use automatic mixed precision to run at least 100 inferences to calculate the averaged data with a warm-up period ahead.
For the FPGA and Edge TPU, we use the results in the original paper. 
The results show that our \name consistently achieves the lowest energy consumption, with over 300 $\times$, 6.6$\times$, 18$\times$, and 20$\times$ reduction compared to CPU, GPU, Edge TPU, and other domain-specific Transformer accelerators. 
For throughput, \name achieves the highest among all the platforms, even on the 4-tile \sname-B system.
We get 2 to 3 orders of magnitude lower energy-delay products for various benchmarks on our \sname-L system.

\subsection{Accuracy \& Robustness Analysis}

We further show that \name can realize digital-comparable accuracy with superior robustness against various non-ideality effects.
We evaluate the accuracy of quantized models running on \name by using the noisy analytic transformation in Eq.~\eqref{eq:power_pair_sub_noise_wdm} during inference.
As shown in Figure~\ref{fig:dispersion_accuracy}, compared to the accuracy of Transformers with equivalent bit-widths running on GPU,
we maintain $<$1\% accuracy loss ensured by our superior robustness.
We also evaluate the robustness of our design against encoding the magnitude noise, phase noise, and dispersion effect.
Figure~\ref{fig:dispersion_accuracy} shows that our design is very robust to WDM dispersion even with more than 20 wavelengths, showing $<$0.5\% accuracy drop.
The reason is our design points of the directional coupler and phase shifter are at the local optima with minimized sensitivity to dispersion-induced perturbations, as discussed in Sec.~\ref{subsec:robustness}.
Figure~\ref{fig:noise_accuracy} shows that our design demonstrates high tolerance to random encoding magnitude and phase variations.
On the DeiT-T ImagenetNet workload, the noise-induced accuracy degradation is within 0.5\% with a small variance
Noise-aware training is adopted in this paper, while
more advanced noise-mitigation techniques~\cite{NP_DATE2020_Gu, HWA_Nature2022_Wang} can be applied to further boost the accuracy and robustness.

%% file: tablels/photonic_device_area.tex
\begin{table}[]
\centering
\small{
\caption{Adopted component parameters in our paper.
IL represents insertion loss, and FSR means free spectral range.
}
\label{tab:device}}
\resizebox{0.9\columnwidth}{!}{
\begin{tabular}{|c|c|c|}
\hline
Device                                    & Parameter   & Value                       \\ \hline
\multirow{3}{*}{DAC~\cite{NP_VLSI2020_Caragiulo}} & Precision     & 8-bit                \\
                     & Power         & 50 mW(@14 GSPS)    \\ 
                     & Area         & 11,000 $\mu$m$^2$   \\\hline
\multirow{3}{*}{ADC~\cite{NP_ISSCC2022_Liu}} & Precision     & 8-bit                \\
                     & Power         & 14.8 mW(@10 GSPS) \\ 
                     & Area         & 2,850 $\mu$m$^2$ \\ 
                     \hline
\multirow{2}{*}{TIA~\cite{NP_VLSI2018_Rakowski}} & Power        & 3 mW                       \\
                                          & Area        & $<$50 $\mu$m$^2$                      \\ \hline
\multirow{4}{*}{Microdisk~\cite{NP_NatureComm2014_Timurdogan}}                & Locking Power        & 0.275 mW\footnote{Scale this to 10nm technology node according to~\cite{NP_NatureComm2014_Timurdogan}} \\
                                          & IL        & 0.93 dB                    \\
                                          & Area        & 4.8 $\times$4.8$\mu$m$^2$  \\
                                          & FSR  & 5.6 THz (55.1 nm) \\ \hline
\multirow{4}{*}{\begin{tabular}[c]{@{}l@{}}Microring \\ Resonator\end{tabular}}      & Tuning Power        & 0.21 mW  \\
                                          & Locking Power        & 1.2 mW/0.5FSR~\cite{NP_OE2013_Streshinisky}  \\
                                          & IL        & 0.95 dB~\cite{NP_LPR2019_Pintus}                      \\
                                          & Area        & 9.66$\times$9.66 $\mu$m$^2$~\cite{NP_LPR2019_Pintus}                          \\ \hline
\multirow{3}{*}{MZM}                      & Tuning Power        & 2.25 mW~\cite{NP_OE2010_Dong}  \\
                                          & IL        & 1.2 dB~\cite{NP_OE2012_Suguru}                      \\
                                          & Area        & 260$\times$20 $\mu$m$^2$~\cite{NP_OE2012_Suguru} \\ \hline
\multirow{2}{*}{\begin{tabular}[c]{@{}l@{}}Directional \\Coupler~\cite{NP_JLT2020_Ye}\end{tabular}} & IL        & 0.33 dB                        \\
                                          & Area        & 5.25$\times$2.4 $\mu$m$^2$                      \\ \hline
\multirow{3}{*}{\begin{tabular}[c]{@{}l@{}}MEMS Phase \\Shifter~\cite{NP_MN2023_Quack}\end{tabular}}            & IL        & 0.33 dB                         \\
                                          & Area        & 100$\times$45 $\mu$m$^2$                     \\ 
                                           & Response Time        & 2$\mu s$                      \\ \hline
\multirow{3}{*}{Photodetetcor~\cite{NP_optical2016_Huang}}             & Power & 1.1 mW\\
                                          & Sensitivity & -25 dBm                      \\
                                          & Area        & 4$\times$10 $\mu$m$^2$                          \\ \hline
\multirow{2}{*}{Y-branch~\cite{NP_IPJ2021_Nair}}                 & IL        & 0.3 dB                        \\
                                          & Area        & 1.8$\times$1.3 $\mu$m$^2$                    \\ \hline
Micro-comb~\cite{NP_Nature2021_Xu} & Area & 1,184$\times$1,184 $\mu$m$^2$ \\\hline
\multirow{2}{*}{On-chip Laser}                 & Wall-plug Efficiency        & 0.2~\cite{NP_CLEO2019_Hao}                        \\
                                          & Area        & 400$\times$300 $\mu$m$^2$                    \\ \hline
\end{tabular}
}
\vspace{-5pt}
\end{table}

%% file: fig_tex/lumerical.tex
\begin{figure}
    \centering
\includegraphics[width=0.4\textwidth]{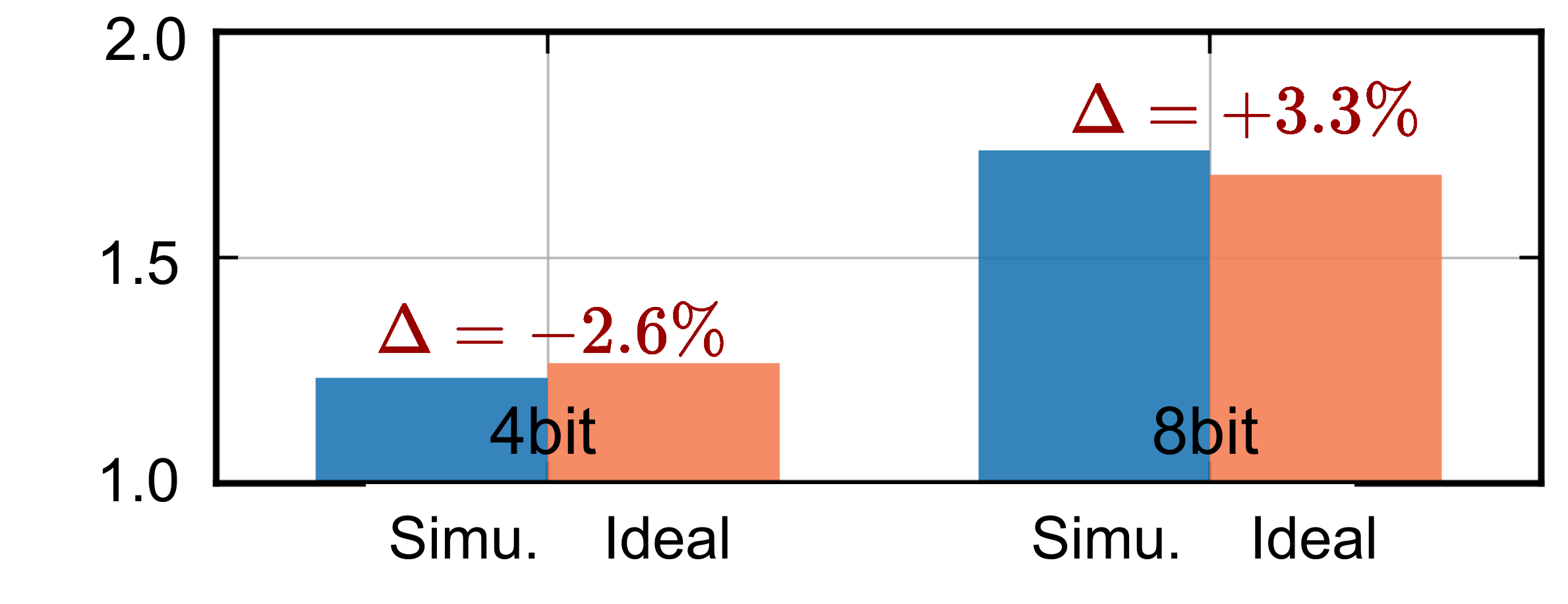}
    \vspace{-5pt}
    \small{\caption{~
    Optical simulation results of 4-bit ad 8-bit random length-12 dot-products on our \oDot engine.
    }
    \label{fig:lumerical}}
    \vspace{-5pt}
\end{figure}

%% file: tablels/dota_params.tex
\begin{table}[t!p]
\centering
\caption{Base (B) and Large (L) configurations for \name.
}
\label{tab:dota_config}
\resizebox{0.93\columnwidth}{!}{
\begin{tabular}{c|cccccc|c}
\hline
Ours & $N_t$ & $N_c$ & $N_h$ & $N_v$ & $N_\lambda$ &\begin{tabular}[c]{@{}c@{}}Global\\SRAM (MB)\end{tabular}  & \begin{tabular}[c]{@{}c@{}}area \\(mm$^2$)\end{tabular}                    \\ \hline
\sname-B   & 4  & 2  & 12 & 12 & 12  &  2 & 60.3                    \\
\sname-L    & 8  & 2  & 12 & 12  & 12 & 4 & 112.82                     \\ \hline
\end{tabular}
}
\vspace{-10pt}
\end{table}

%% file: fig_tex/system_area.tex
\begin{figure}[p!t]
    \centering
    \vspace{-5pt}
    \subfloat[]{
        \includegraphics[width=0.23\textwidth]{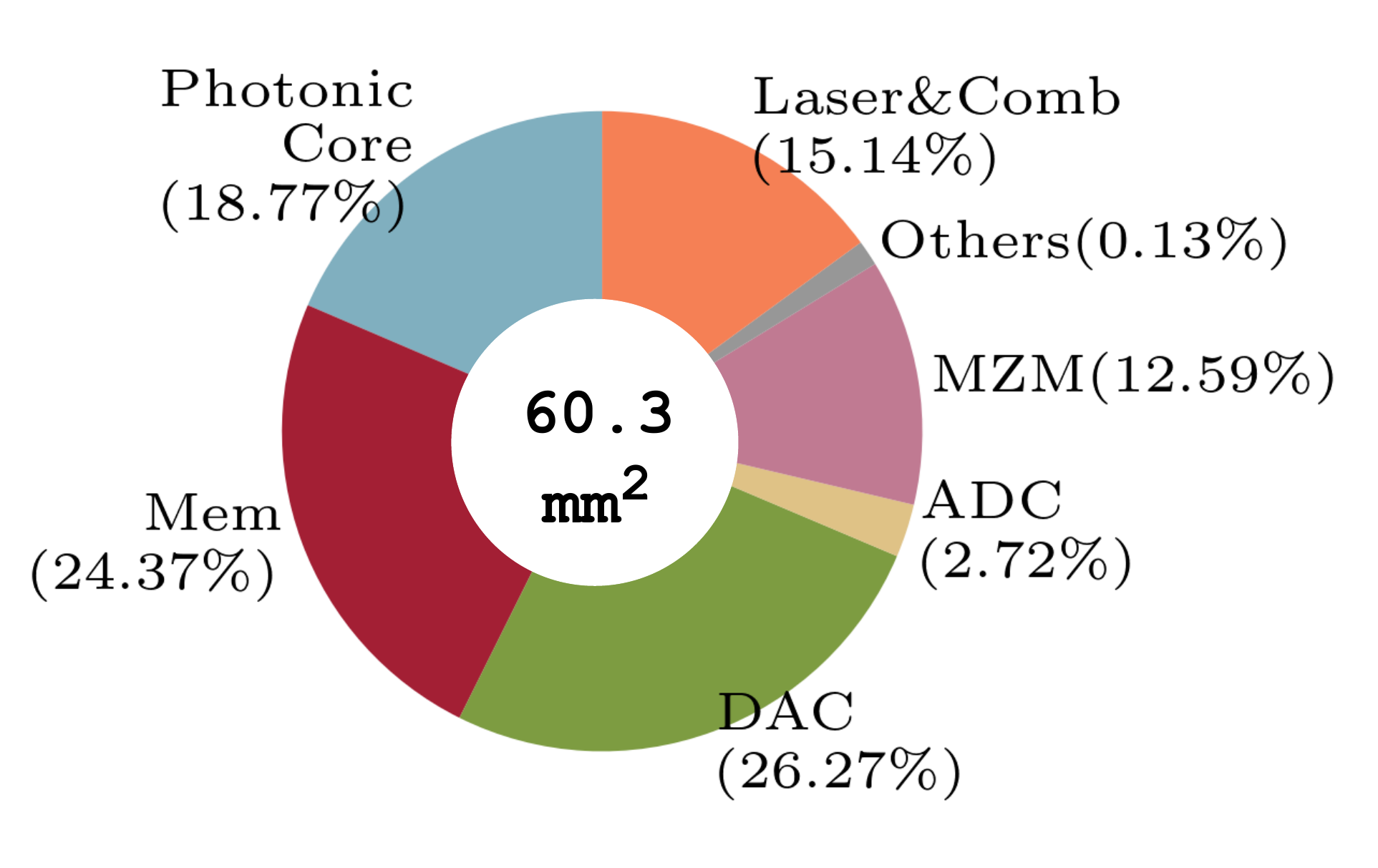}
        \label{fig:4bitDota}
    }
    \subfloat[]{
        \includegraphics[width=0.23\textwidth]{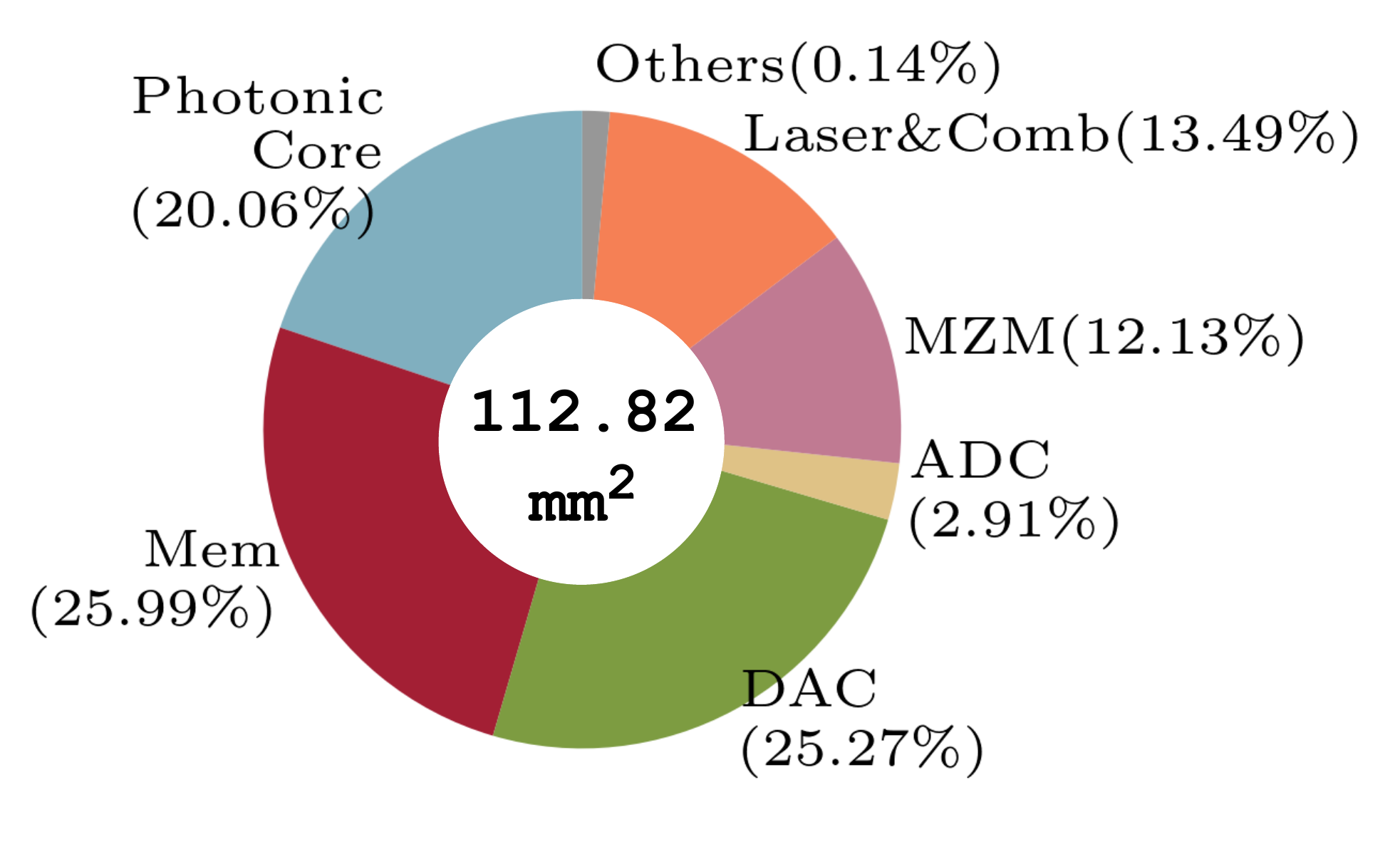}
        \label{fig:8bitDota}
    }
    \caption{~
    Area breakdown of (a) \sname-B and (b) \sname-L.
    }
    \label{fig:DotaArea}
    \vspace{-15pt}
\end{figure}

%% file: fig_tex/system_power.tex
\begin{figure}[p!t]
    \centering
    \vspace{-5pt}
    \subfloat[]{
        \includegraphics[width=0.22\textwidth]{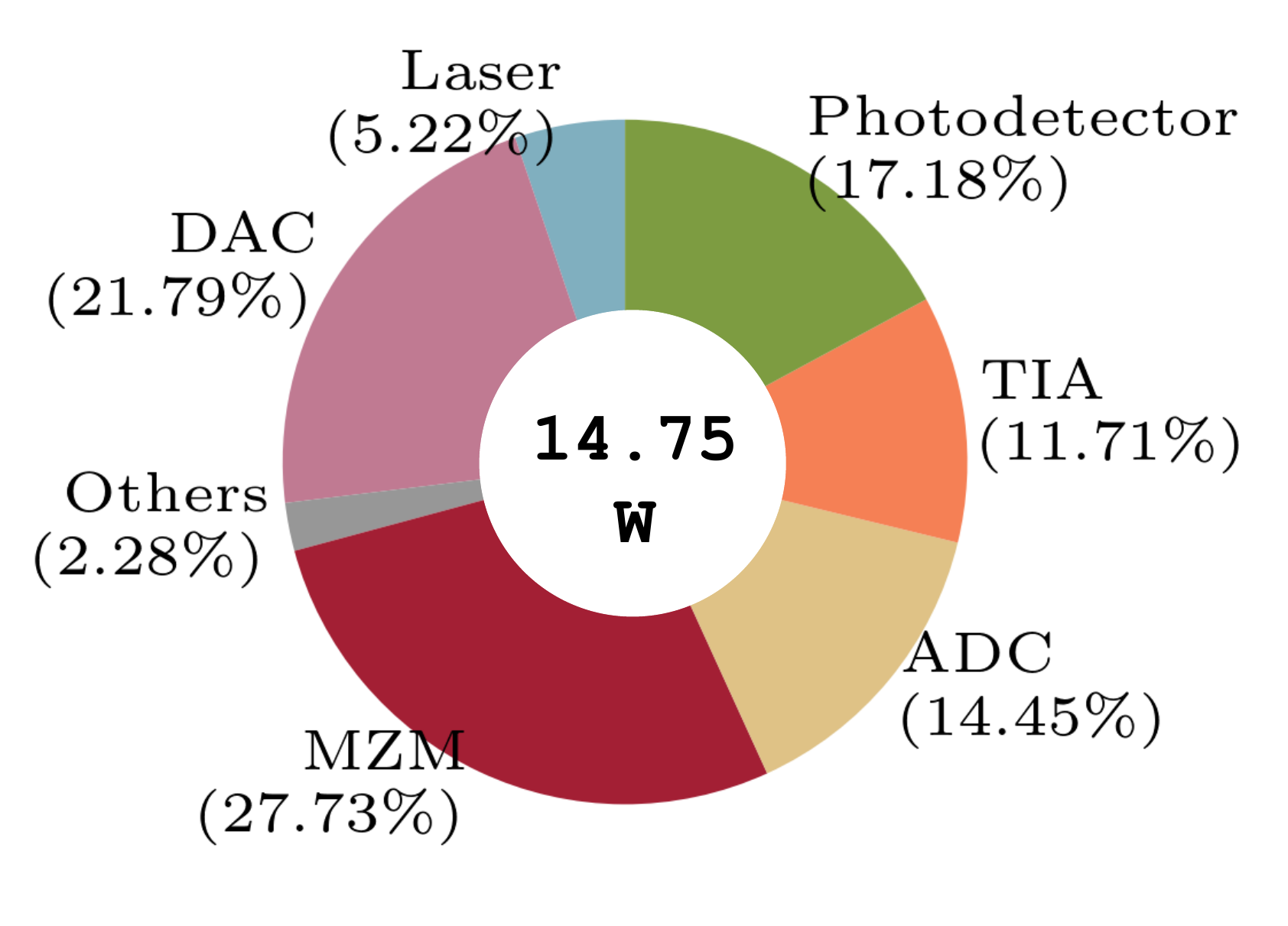}
        \label{fig:4bitDota_power}
    }
    \subfloat[]{
        \includegraphics[width=0.22\textwidth]{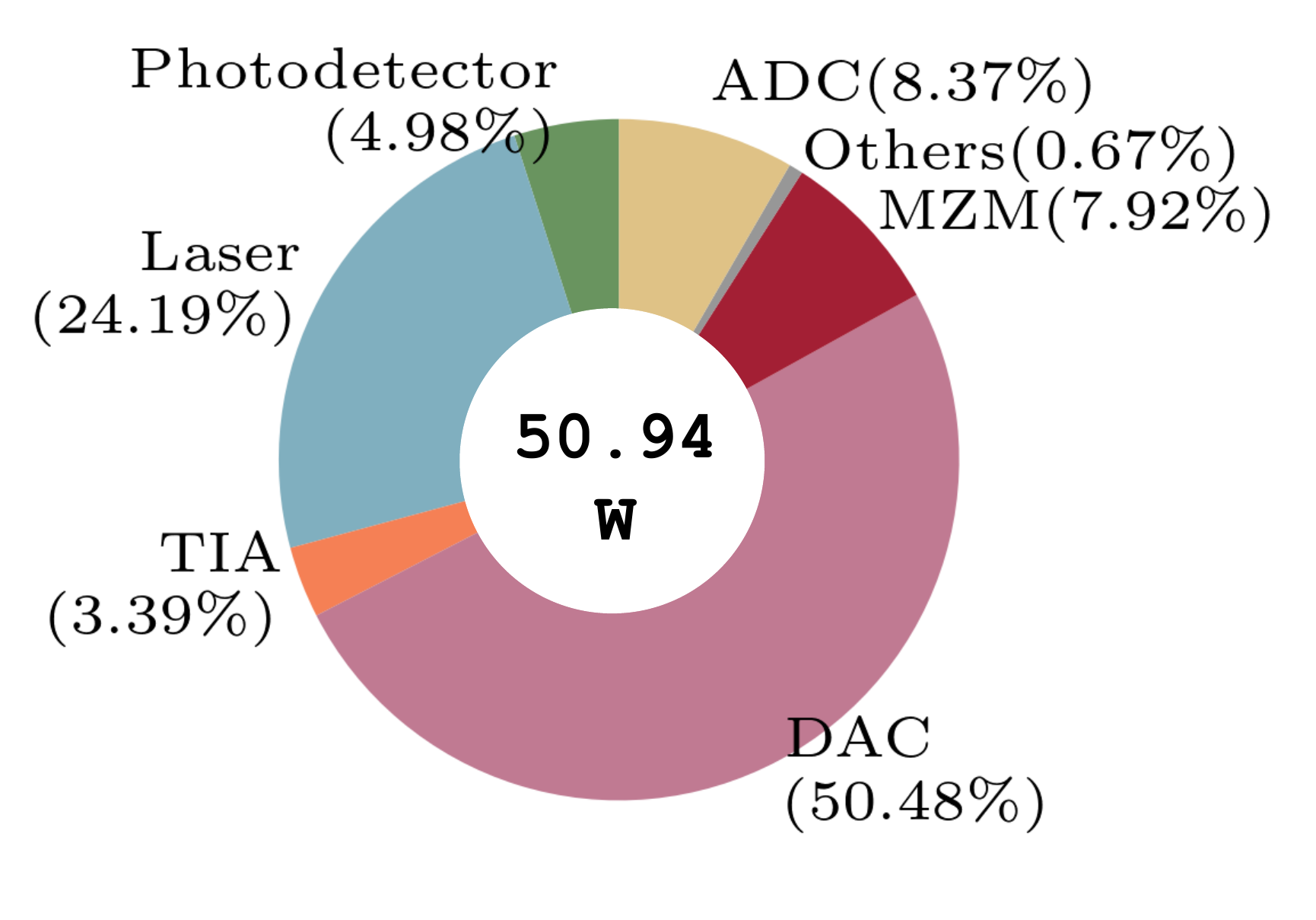}
        \label{fig:8bitDota_power}
    }
    \vspace{-5pt}
    \caption{~
    Power breakdown of \sname-B for (a) 4-bit (b) 8-bit precision.  Others include memory system static power.
    }
    \label{fig:system_power}
    \vspace{-10pt}
\end{figure}

%% file: fig_tex/fig_ptc_scaling.tex
\begin{figure*}[p!t]
    \centering
\includegraphics[width=0.95
\textwidth]{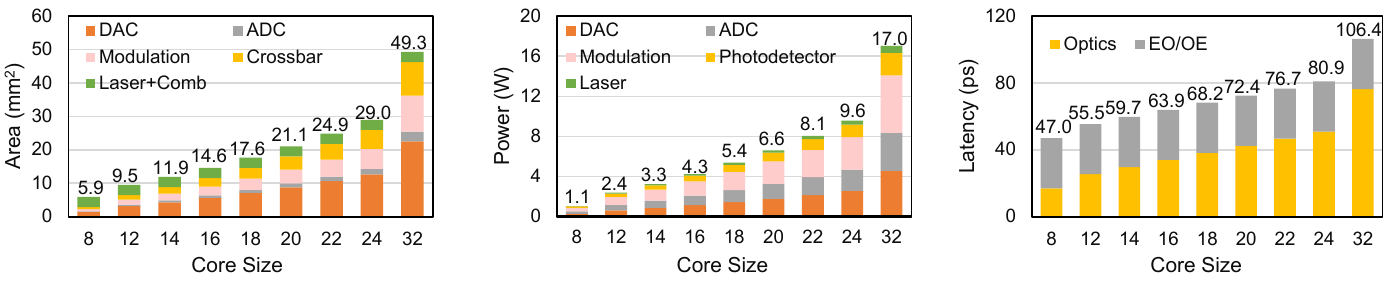}
\label{fig:ptc_area_power_latency_scale}
\vspace{-3pt}
    \caption{~Area, power, and latency scaling of single 4-bit \oMM core with increasing core size $N$, where $N_h = N_w = N_{\lambda} = N$.
    }
    \label{fig:ptc_scaling}
\end{figure*}

%% file: fig_tex/fig_ptc_perf_scale.tex
\begin{figure}[t!]
    \centering
\includegraphics[width=0.9\columnwidth]{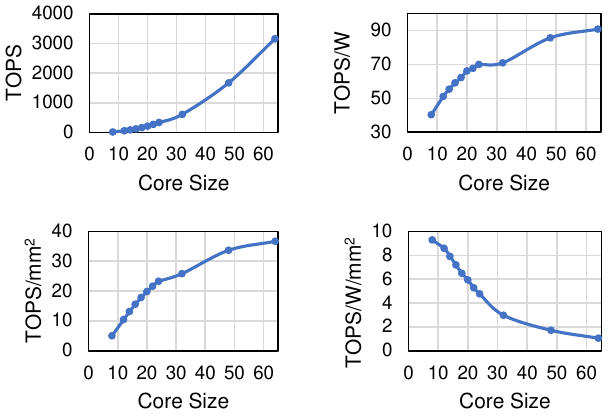}
\vspace{-5pt}
    \caption{~Performance and efficiency of the optical computing part (ADC/DAC excluded) scale with larger core size.}
    \vspace{-5pt}
    \label{fig:ptc_perf_scale}
\end{figure}

%% file: fig_tex/fig_attention_linear.tex
\begin{figure}[t!]
    \centering
\includegraphics[width=0.8\columnwidth]{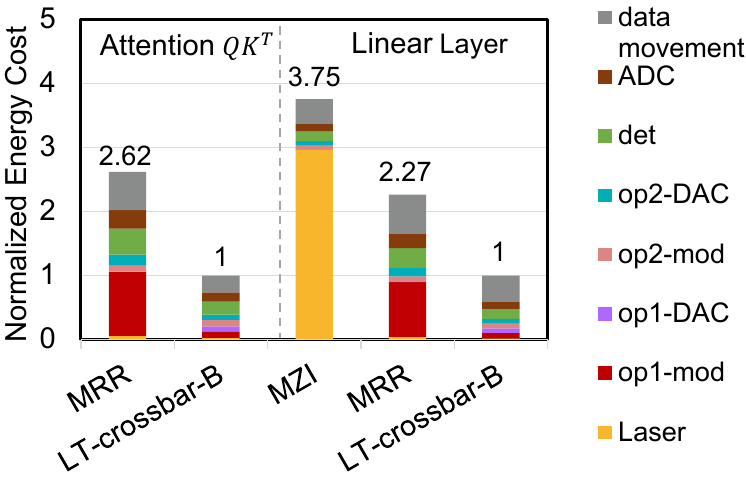}
    \caption{~Energy comparison and breakdown across main components (adder excluded) across \sname-\texttt{crossbar}-B (\sname-B without architecture-level optimization), MRR bank, and MZI array on example attention and linear layer workloads in DeiT-T. \emph{mod} refers to the energy cost of dynamic modulation and operand locking. \emph{det} refers to the energy costs of Photodetector and TIA. op1 is the static operand in MRR and MZI baselines when using weight-static dataflow. 
    }
    \vspace{-10pt}
    \label{fig:compare_linear_attention}
\end{figure}

%% file: tablels/compare_prior_onn.tex
\begin{table*}[]
\centering
\caption{Comparison to prior photonic accelerators on in Deit-T/B.
The MHA ($Qk^T$ and $AV$) and FFN modules are highlighted.
We assume MRR bank implements MHA in the MZI array as it cannot support MHA.
Our \name is equipped with our architecture-level optimization discussed in~Section~\ref{subsec:opt}, while even without, still better than baselines.
}
\label{tab:compare_prior_onns}
\resizebox{0.96\textwidth}{!}{
\begin{tabular}{|ccc|ccc|ccc|
>{\columncolor[HTML]{EFEFEF}}c 
>{\columncolor[HTML]{EFEFEF}}c 
>{\columncolor[HTML]{EFEFEF}}c 
>{\columncolor[HTML]{EFEFEF}}c |}
\hline
                                             &                                               &        & \multicolumn{3}{c|}{MZI-based Accelerator~\cite{NP_NATURE2017_Shen}}                                                                                                                                               & \multicolumn{3}{c|}{MRR-based Accelerator~\cite{NP_SciRep2017_Tait}}                                                                                                                                                                & \multicolumn{4}{c|}{\cellcolor[HTML]{EFEFEF}\name-B}                                                                                                                      \\ \hline
\multicolumn{1}{|c|}{Precision}             & \multicolumn{1}{c|}{Model}                   & Module & \begin{tabular}[c]{@{}c@{}}Energy$\downarrow$\\ (mJ)\end{tabular} & \begin{tabular}[c]{@{}c@{}}Latency$\downarrow$\\ (ms)\end{tabular} & \begin{tabular}[c]{@{}c@{}}EDP$\downarrow$\\ (mJ*ms)\end{tabular} & \begin{tabular}[c]{@{}c@{}}Energy$\downarrow$\\ (mJ)\end{tabular} & \begin{tabular}[c]{@{}c@{}}Latency$\downarrow$\\ (ms)\end{tabular} & \begin{tabular}[c]{@{}c@{}}EDP$\downarrow$\\ (mJ*ms)\end{tabular} & \begin{tabular}[c]{@{}c@{}}Energy w/o\\ Arch Opt $\downarrow$ (mJ)\end{tabular} & \begin{tabular}[c]{@{}c@{}}Energy $\downarrow$ \\ (mJ)\end{tabular} & \begin{tabular}[c]{@{}c@{}}Latency$\downarrow$\\ (ms)\end{tabular} & \begin{tabular}[c]{@{}c@{}}EDP$\downarrow$\\ (mJ*ms)\end{tabular} \\ \hline
\multicolumn{1}{|c|}{\multirow{6}{*}{4bit}} & \multicolumn{1}{c|}{\multirow{3}{*}{Deit-T}} & MHA    & -           & -           & -           & 0.17  & 0.03 & 0.005  & 0.08  & 0.04  & 3.12e-3 & 1.33e-4 \\
\multicolumn{1}{|c|}{}  &   \multicolumn{1}{c|}{}     & FFN & 1.47   & 6.27   & 9.19      & 0.89  & 0.14 & 0.12  & 0.39  & 0.22  & 1.04e-2 & 2.28e-3  \\
\multicolumn{1}{|c|}{}  &   \multicolumn{1}{c|}{}     & All & 2.98   & 12.37  & 36.93     & 1.54  & 0.24 & 0.38  & 0.69  & 0.38  & 1.94e-2 & 7.44e-3  \\ \cline{2-12} 
\multicolumn{1}{|c|}{}                      & \multicolumn{1}{c|}{\multirow{3}{*}{Deit-B}} & MHA & -           & -           & -            & 0.67  & 0.12 & 0.08  & 0.34  & 0.17  & 1.25e-2 & 2.13e-3        \\
\multicolumn{1}{|c|}{}                      & \multicolumn{1}{c|}{}        & FFN & 23.46  & 100.24 & 2351.23   & 14.16 & 2.21 & 31.34 & 6.25  & 3.47  & 1.67e-1 & 5.81e-1       \\
\multicolumn{1}{|c|}{}                      & \multicolumn{1}{c|}{}       & All & 44.91  & 190.46 & 8554.08   & 22.08 & 3.47 & 76.56 & 9.79  & 5.44  & 2.65e-1 & 1.44         \\
                       \hline
\multicolumn{3}{|c|}{\textbf{Average Ratio}}                                                        & \textbf{8.01}                                        & \textbf{677.56}                                        & \textbf{5426.27}                                      & \textbf{4.03}                                      & \textbf{12.85}                                       & \textbf{51.79}                                                     & \textbf{1.80}   & \textbf{1}                                          & \textbf{1}                                              & \textbf{1}                                            \\ \hline
\hline
\multicolumn{1}{|c|}{\multirow{6}{*}{8bit}} & \multicolumn{1}{c|}{\multirow{3}{*}{Deit-T}} & MHA     & -            & -           & -           & 0.36  & 0.03 & 0.01  & 0.25  & 0.15  & 3.12e-3 & 4.76e-4        \\
\multicolumn{1}{|c|}{}                      & \multicolumn{1}{c|}{}                        & FFN    & 19.21  & 6.27   & 120.32    & 1.83  & 0.14 & 0.25  & 1.09  & 0.68  & 1.04e-2 & 7.08e-3        \\
\multicolumn{1}{|c|}{}                      & \multicolumn{1}{c|}{}                        & All & 37.18  & 12.37  & 460.09    & 3.20  & 0.24 & 0.78  & 1.93  & 1.21  & 1.94e-2 & 2.34e-2       \\ \cline{2-12} 
\multicolumn{1}{|c|}{}                      & \multicolumn{1}{c|}{\multirow{3}{*}{Deit-B}} & MHA & -            & -           & -    & 1.43  & 0.12 & 0.12  & 1.02  & 0.61  & 1.25e-2 & 7.61e-3        \\
\multicolumn{1}{|c|}{}                      & \multicolumn{1}{c|}{}                        & FFN    & 307.27 & 100.24 & 30800.44  & 29.33 & 2.21 & 2.21  & 17.40 & 10.81 & 1.67e-1 & 1.81        \\
\multicolumn{1}{|c|}{}                      & \multicolumn{1}{c|}{}                        & All    & 580.80 & 190.46 & 110620.44 & 45.77 & 3.47 & 3.47  & 27.33 & 16.98 & 2.66e-1 & 4.51   \\ \hline
\multicolumn{3}{|c|}{\textbf{Average Ratio}}                                                        & \textbf{32.46}                                        & \textbf{675.67}                                        & \textbf{21944.30}                                     & \textbf{2.67}                                       & \textbf{12.81}                                       & \textbf{34.25}                                                  & \textbf{1.61}    & \textbf{1}                                          & \textbf{1}                                              & \textbf{1}                                            \\ \hline
\end{tabular}  
}
\end{table*}

%% file: fig_tex/fig_detail_breakdown_compare_mrr.tex
\begin{figure}[t!]
    \centering
\includegraphics[width=0.9\columnwidth]{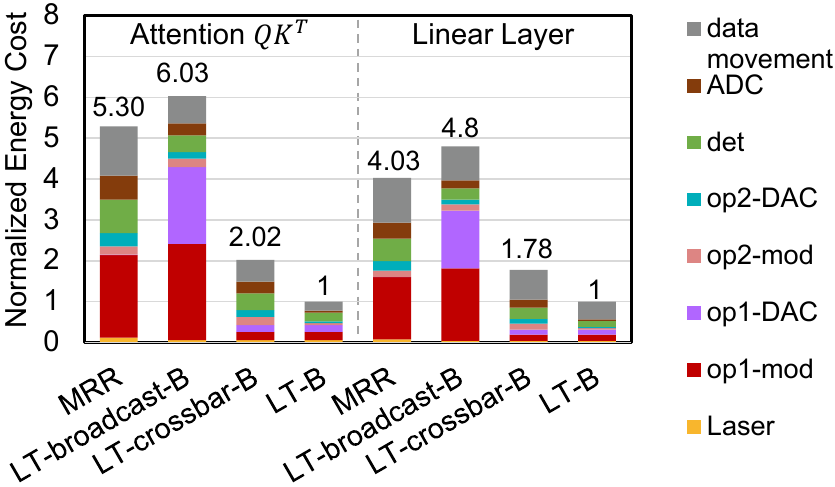}
    \caption{~Energy comparison and breakdown across main components between MRR and \sname-B variants on example workloads ($QK^T$ and first layer of FFN) in DeiT-T. op1 is the static operand when MRR uses weight-static dataflow.
    \sname-\texttt{broadcast}-B
    adopts a simple topology to only broadcast input operand (op1).
    \sname-\texttt{crossbar}-B is our \oMM design that employs the crossbar topology to
    share both operands intra-core.
    \sname is complete version that furthers adopts architecture-level optimization.}
    \vspace{-10pt}
    \label{fig:compare_linear_attentio_breakdown}
\end{figure}

%% file: fig_tex/fig_compare_energy_fps.tex
\begin{figure*}[p!t]
    \centering
        \includegraphics[width=0.95\textwidth]{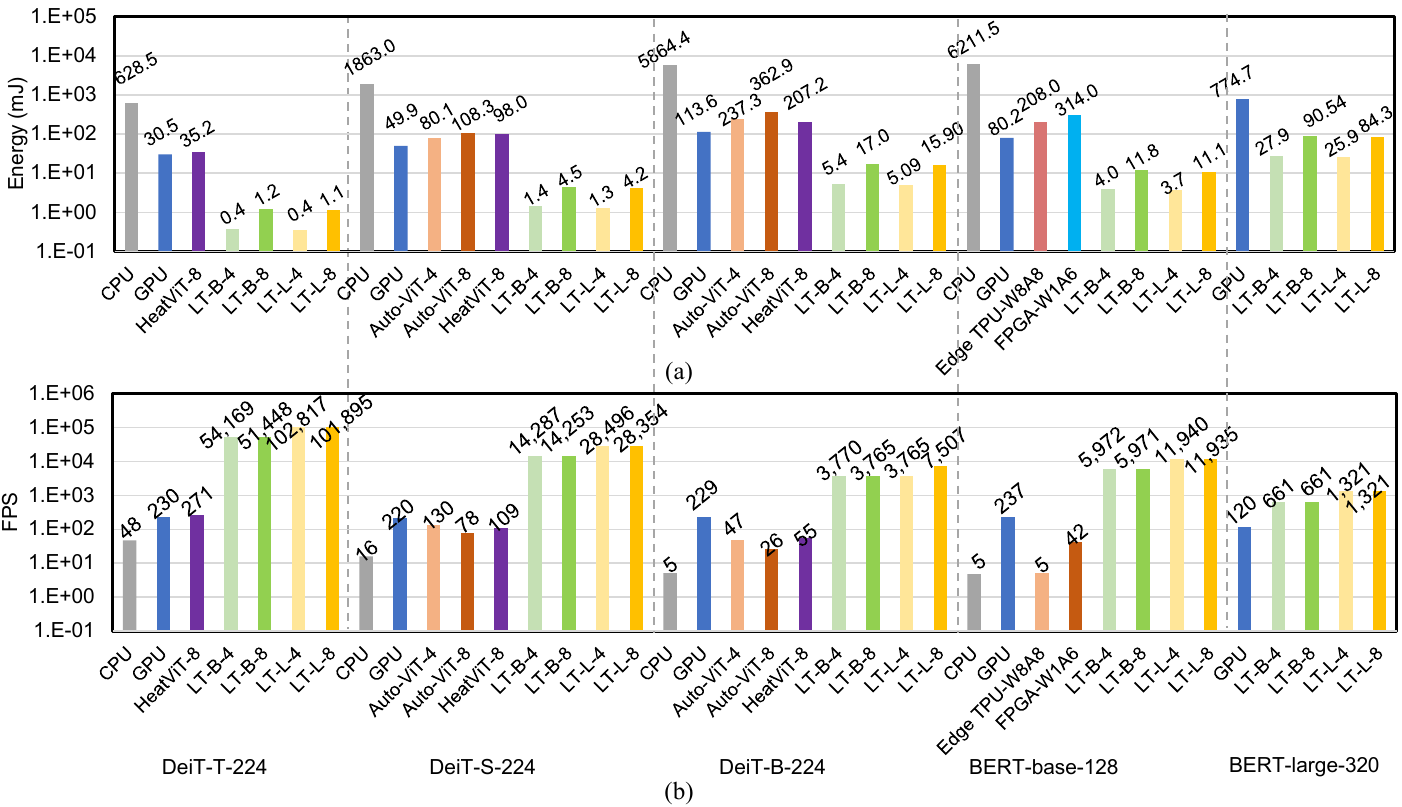}
        \vspace{-5pt}
    \caption{Compare (a) energy consumption (mJ) and (b) frames-per-second (FPS) across different accelerator designs on various workloads (DeiT-series with ImageNet1K-224x224 and BERT-series with 128 and 320 sequence lengths) and two different bitwidth (4-bit and 8-bit).  
    }
    \label{fig:compare_energy_fps}
    \vspace{-10pt}
\end{figure*}

%% file: fig_tex/fig_dispersion_accuracy.tex
\begin{figure}[p!t]
    \centering
    \includegraphics[width=0.9\columnwidth]{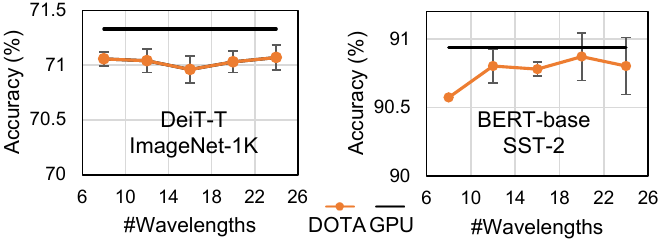}
    \caption{~Dispersion robustness evaluation with a range of wavelengths on 4-bit DeiT-T ImageNet1K and 8-bit BERT-base SST2 workload.
    The accuracy of quantized models running on GPU without any noise is shown. Input noise std. and phase noise std. are set to 0.03 and 2$^\circ$.
    }
\label{fig:dispersion_accuracy}
\vspace{-10pt}
\end{figure}

%% file: fig_tex/fig_noise_accuracy.tex
\begin{figure}[t!]
    \centering
\includegraphics[width=0.9\columnwidth]{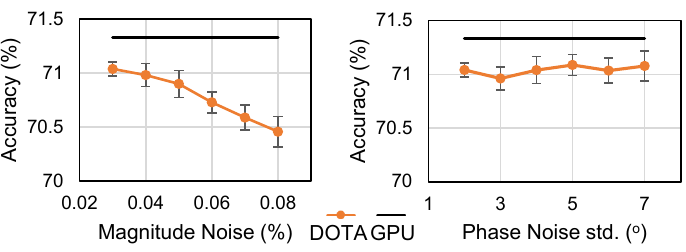}
    \caption{~Encoding magnitude and phase noise robustness evaluation with varying noise intensity on 4-bit DeiT-T ImageNet1K.
    }
    \label{fig:noise_accuracy}
    \vspace{-10pt}
\end{figure}

%% file: doc/discuss.tex
\section{Discussion}
\label{sec:discuss}

\subsection{Sparsity Support}

\input{fig_tex/sparse_support}

The sparsity opportunity in Transformers mainly lies in three folds~\cite{HWA_HPCA2023_Dong}.
(1) \textbf{Redundancy of the attention head/token}. Attention head/token pruning directly removes unimportant heads or tokens entirely~\cite{HWA_HPCA2021_Wang}.
(2) \textbf{Redundancy of the token channels}. Token channel pruning removes some channels of the token embedding.
(3) \textbf{Redundancy of the attention map}. A number of Transformers have proposed to explore sparsity inside the attention map so as to make the attention module more efficient. For example, BigBird~\cite{NN_Neurips2020_Zaheer} and BlockBERT~\cite{NN_EMNLP2020_Qiu} propose structured/block-wise sparse attention patterns, e.g., window- and global- attentions. The sparse attention features sparse computation in both $QK^T$ and $AV$.

Our \name can be easily extended to support the (1) and (2) opportunities, as they remove head/token/channel entirely, resulting in regular dense GEMM. 
We can also support hardware-friendly structured/block-wise (3) sparse attention patterns by reformulating sparse computation into small chunked dense matrix-matrix/vector-matrix multiplication, as illustrated in Figure~\ref{fig:sparse_support}.
To generate sparse attention $A=QK^T$ efficiently, we can blockify Q/K matrices based on structured sparse patterns and form groups of matrix-matrix multiplication.
Take the block-wise window local attention~\cite{NN_Neurips2020_Zaheer} as an example.
Assume the number of tokens is $n$, the window size is $w$ and the block size is $b$. 
The token $i$ will only attends to key matrix $K$ with index $i-(w-1)/2$ to $i+(w-1)/2$.
We can blockify the $Q$ and $K$ matrix based on the block size $b$, resulting in $\lceil n/b\rceil$ chunked $Q$ and $K$ matrices. 
Based on the window pattern, each chunked $Q$ matrices will compute with only $w$ chunked $K$ matrices, still featuring dense matrix-matrix multiplication.
To support $AV$, we can first compress the sparse attention $A$ row by row. This approach generates dense $A$ chunked matrices/vectors based on whether the sparsity is block-wise or not. 
Then, we can form vector-matrix/matrix-matrix multiplication between the compressed $A$ with the corresponding rows of $V$.
The above reformulated dense matrix-matrix/matrix-vector multiplication can be efficiently accelerated by \oMM. 
Moreover, we have the flexibility to explore heterogeneous \texttt{DPTC}s by having different/searched core sizes in Table~\ref{tab:arch_params} to better suit workloads with specific sparse patterns, avoiding low-utilization scenarios.
For example, we can have a specific \oMM engine for vector-matrix multiplication by setting $N_h$ to 1 to support vector-matrix multiplication featured by non-block-wise sparsity.

\subsection{Large Language Model Support}

Supporting large language models (LLMs) on our accelerator presents both challenges and opportunities, primarily because LLMs rely on parallel-unfriendly autoregressive Transformer models. 
These models generate tokens \textit{one at a time} based on input and previous tokens, resulting in small-dimensional matrix multiplications with low operation intensity. This characteristic makes LLMs memory-bounded and underutilized the ultra-fast computing power offered by the photonic chips.
Besides, they demand substantial on-chip memory for storing key and value tensors, also known as KV cache.

To adapt our accelerator for LLMs, 
several strategies can be considered. 
First, scaling on-chip memory or even building multi-chip systems are necessary to meet the memory/throughput demands of LLMs adequately. 
Secondly, we can batch multiple requests together to increase operation intensity, making better use of the accelerator's computing capabilities. 
Additionally, we can trade excessive memory demands for cost-effective and rapid optical computation by recalculating Query (Q) and Key (K) values instead of fully caching them, as demonstrated in recent work~\cite{xiao2023efficient}. 
Model compression techniques~\cite{HWA_HPCA2021_Wang, zhang2023h} can also be applied to reduce memory usage by pruning unimportant tokens.
Furthermore, optimizing our tiling algorithm, similar to the approach in~\cite{dao2022flashattention}, can help avoid materializing large full attention matrices, reducing reliance on slow off-chip DRAM and improving overall efficiency. 

By implementing these strategies and further tailoring our accelerator,
we believe our work is poised to extend its
capabilities to support Transformer-backboned large language models (LLMs) in the future.

%% file: fig_tex/sparse_support.tex
\begin{figure}
    \centering
\includegraphics[width=0.48\textwidth]{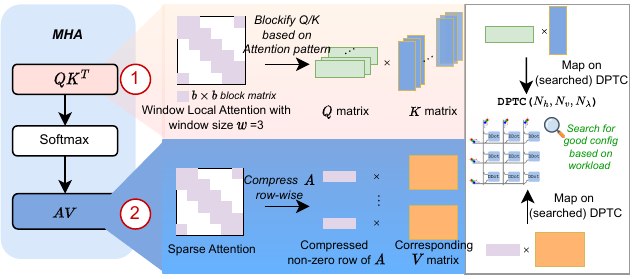}
    \small{\caption{
    Illustration of sparse Attention support on \name using window-based local attention as an example. After blockification/compression, we transform the sparse computation to dense matrix/vector-matrix multiplication that \oMM can accelerate efficiently.
    }
    \label{fig:sparse_support}}
    \vspace{-10pt}
\end{figure}

%% file: doc/7conclu.tex
\vspace{-.05in}
\section{Conclusion}
\label{sec:Conclusion}
The proliferation and increasing complexity of attention-based Transformers have spurred the need for specialized hardware accelerators.
Photonic accelerators have shown promising efficiency and speed for CNN workloads, which require only weight-static linear operations. 
However, the current SoTA photonic accelerators face challenges in supporting Transformer with self-attention operations, primarily due to their inability to handle dynamic tensor multiplication and encode full-range operands.
In this work, we introduce the first customized, high-performance, and energy-efficient photonic Transformer accelerator, \name. 
We propose a novel photonic tensor core, a crossbar array of dynamic optical vector dot-product engines. 
This design overcomes the fundamental constraints of existing designs, enabling ultra-parallel matrix multiplication of two dynamic, full-range matrices. 
Our comprehensive evaluation shows that \name achieves over 2.6$\times$ energy and over 12$\times$ latency reductions compared to prior photonic accelerators. 
Furthermore, it outperforms electronic Transformer accelerators with over 6$\times$ energy reduction and 2 to 3 orders of magnitude lower energy-delay products with digital-comparable accuracy.
Looking ahead, our work is poised to extend its capabilities to support Transformer-backboned large language models (LLMs) in the future, further enhancing its applicability to cutting-edge machine learning tasks.

Our work highlights the potential of domain-specific photonic AI hardware for the efficient acceleration of advanced ML tasks.
In the future, we anticipate significant advancements in optical computing technologies, enhancing their flexibility, applicability, and performance across a broader range of machine learning tasks. 
Ongoing research in innovative cross-layer co-design will push the boundaries of what photonic accelerators can achieve and result in the creation of highly efficient, low-latency next-generation electronic-photonic computing systems for increasingly complex ML workloads.

%% file: doc/8acknowledge.tex
\section*{Acknowledgment}
This work is supported in part by the Multidisciplinary University Research Initiative (MURI) program through the Air Force Office of Scientific Research (AFOSR) under contract \#FA 9550-17-1-0071 and AFOSR project \#FA9550-23-1-0452. 
We thank all anonymous HPCA reviewers for their insightful comments.
Thanks to Zhixing Jiang and Shupeng Ning from the University of Texas at Austin for helpful suggestions and help during the artifact evaluation process.

%% file: doc/appendix.tex
\appendix
\section{Artifact Appendix}

\subsection{Abstract}
Lightening-Transformer introduces the first customized photonic Transformer accelerator built upon a novel dynamically-operated photonic tensor core design. 

The artifact contains three parts.
The first is the noise-aware training/inference framework, enabling users to train/infer with the optical Transformer models built onto our photonic tensor core. We embed the analytic transformation of PTC in the forward path when computing matrix multiplication.
We support quantization and non-ideality injection, including input phase/magnitude variation, WDM-induced dispersion, and a general systematic error term.
The second part is the hardware simulator for our \name accelerator.
Our simulator simulates performance using behavior-level models.   
Our simulator can estimate the area and power of our \name accelerator.
Besides, it can predict the energy and latency when running Transformer workloads (DeiT/BERT) on \name.
The third part is profiling scripts for measuring the latency and power when running workloads on GPUs, such that you can compare the GPU performance with our light-empowered accelerators.
We provide scripts for each part to validate our performance.

The minimal hardware requirement will be one CPU and one Nvidia GPU.
The minimal software requirements will be CUDA and Python libraries such as PyTorch.

\subsection{Artifact check-list (meta-information)}

{\small
\begin{itemize}
  \item {\bf Program: } Python
  \item {\bf Model: } Low-bit optical Transformer models.
  \item {\bf Data set: } Publicly available Image classification dataset for DeiT evaluation. 
    \begin{itemize}
        \item ImageNet: \href{http://image-net.org/}{http://image-net.org/}: 160 GB.
    \end{itemize}
    The dataset needs extra download and preparation, which is not included in our artifact, given its large size. Download and extract ImageNet using the \href{https://gist.github.com/bonlime/4e0d236cf98cd5b15d977dfa03a63643}{script}. 
  \item {\bf Run-time environment: }
  \begin{itemize}
      \item Ubuntu18 or above
      \item Main software dependencies: Python, PyTorch, pytorch-image-models 0.3.2.
      \item conda is required for package management.
  \end{itemize}
  \item {\bf Hardware: }
  \begin{itemize}
      \item At least one Nvidia GPU for running inference of optical Transformer models with noise injection.
      \item Training with optical Transformer models on ImageNet with quantization and modeled PTC behavior can be time-consuming and GPU-hungry.
       Hence, we provide a trained checkpoint for users to quickly validate accuracy under various non-idealities.  
      \item We tested on Nvidia A100 and A6000 GPUs.
  \end{itemize}
  \item {\bf Output: }
  \begin{itemize}
      \item Classification accuracy of optical Transformer models on ImageNet.
      \item Area and power report of our accelerator.
      \item Latency and energy estimation of our accelerator for given workloads.
  \end{itemize}
  \item {\bf Experiments: }
  We prepare shell scripts for the following experiments.
  \begin{itemize}
      \item Inference of the 4-bit DeiT-T (checkpoint provided) on ImageNet dataset with various variations injected.
      \item Training of low-bit optical DeiT model (Optional).
      \item Area and power profiling for our accelerator \name.
      \item Latency and energy estimation on our accelerator \name for DeiT/BERT workloads.
      \item Workload profiling on GPU.
  \end{itemize}
  \item {\bf How much disk space required (approximately)?: } $<$ 10GB without considering the ImageNet data.
  \item {\bf How much time is needed to prepare workflow (approximately)?: } 1 hour.
  \item {\bf How much time is needed to complete experiments (approximately)?: } 24 hours without training.
  \item {\bf Publicly available?: } Yes. 
  \begin{itemize}
      \item \href{https://github.com/zhuhanqing/Lightening-Transformer}{https://github.com/zhuhanqing/Lightening-Transformer}
      \item \href{https://figshare.com/articles/software/Lightening-Transformer/24899037}{https://doi.org/10.6084/m9.figshare.24899037.v1}
  \end{itemize}
  \item {\bf Code licenses (if publicly available)?: } GNU GPL3.0
  \item {\bf Data licenses (if publicly available)?: }
  \begin{itemize}
      \item ImageNet. BSD 3-Clause License
  \end{itemize}
  \item {\bf Workflow framework used?: } PyTorch.
  \item {\bf Archived (provide DOI)?}
  \href{https://doi.org/10.6084/m9.figshare.24899037.v1}{https://doi.org/10.6084/m9.figshare.24899037.v1}.
\end{itemize}
}

\subsection{Description}

\subsubsection{How to access}
The artifact is available at \href{https://github.com/zhuhanqing/Lightening-Transformer}{Github repo}.

Check the \texttt{readme.md} for the detailed descriptions. Basically, it has three parts 
\begin{itemize}
    \item software\_model: algorithm codes for training/running optical Transformers on our photonic accelerator, which models the matrix multiplication with the analytical transformation of our unique photonic tensor core.
    Quantization and noise injection are supported.
    \item hardware\_simulator: our behavior-level simulator.
    \item profile: profiling scripts for latency and power measurement on GPU.
\end{itemize}

\subsubsection{Hardware dependencies}
Python scripts are deployed on the server with a dedicated Nvidia GPU(CUDA$>$11.x). These Python files are implemented for command run on the server. Our implementations have been evaluated with Nvidia A100 and A6000.
\begin{itemize}
    \item One power GPU machine is desired.
    \item The training of DeiT-T on 4 A100 takes $\sim$2 days on ImageNet due to the quantization operations and dedicated simulation of GEMM workloads on our photonic tensor core.
    \item Hence, we provide a checkpoint that can be evaluated for accuracy validation. It takes 30 mins to obtain the inference accuracy on a single A100, with variation injection and quantization.
\end{itemize}

\subsubsection{Software dependencies}
The artifact is implemented in Python and requires several packages, such as PyTorch and Timm. 
The detailed install step can follow the \texttt{readme.md} in the artifact. 
We use \texttt{conda} to manage packages.

\subsubsection{Data sets}
Image classification datasets ImageNet.  
Download and extract ImageNet following \href{https://github.com/facebookresearch/deit/edit/main/README_deit.md}{facebookresearch/deit}.
This \href{https://gist.github.com/bonlime/4e0d236cf98cd5b15d977dfa03a63643}{script} provides a step-by-step instruction.

\subsubsection{Models}
In this artifact, we provide the codes for the optical DeiT model with quantization and various non-idealities injection.
We support injecting input magnitude encoding variation, input phase encoding variation, output variation, and WDM-induced dispersion, as discussed in Sec.~\ref{subsec:robustness}.

\subsection{Installation}

You can first download the repo to the local machine.
Then, enter the folder and install the required packages following the instructions in the README.md.

\subsection{Experiment workflow}
We provide multiple examples to run our artifact to reproduce the main results of our papers.

\subsubsection{Inference with the optical Transformer model}
Enter the \texttt{software\_model} directory for this part.
We provide a script (\texttt{evaluate\_quant\_transformer\_scan\_noise.sh}) to construct and infer an optical Deit-T model with sweeping noises such that you can obtain the results in Fig.~\ref{fig:dispersion_accuracy} and Fig.~\ref{fig:noise_accuracy}

Check the \texttt{./software\_model/readme.md} for more details if you want to customize experiments, e.g., launching training jobs or inference with a different noise value.

\subsubsection{Area and power profiling of our accelerator}
Enter the \texttt{hardware\_simulator} directory.
\begin{itemize}
    \item Run \texttt{entry\_area\_power\_profile.py} to obtain area and power report.
    \item Results is dumped out in CSV format. It should deliver the same results as in Figure~\ref{fig:DotaArea} and Figure~\ref{fig:system_power}.
\end{itemize}

We provide a batch run script such that you can profile the area and power of all \name-variants. See \texttt{./scripts/area\_power\_all.sh}
All results will be in \texttt{./results/area\_power\_all/}.

\subsubsection{Latency and energy estimation of running workloads in our accelerator}
\begin{itemize}
    \item Run \texttt{./scripts/energy\_latency\_onns\_deit.sh} to compare with photonic baselines.
    \item Run \texttt{./scripts/energy\_latency\_all.sh} to obtain energy and latency  of our accelerator on various DeiT/BERT models.
\end{itemize}

\subsubsection{Profile latency and energy cost of inference on GPU}
Enter the \texttt{profile} directory and follow the \texttt{readme.md} to measure the latency and energy cost for DeiT/BERT models.

\subsection{Evaluation and expected results}
The expected results should match what we have in the experimental session and demonstrate the orders-of-magnitude better energy efficiency of our \name accelerator over baselines.

\subsection{Experiment customization}

This framework provides various customization.
\begin{itemize}
    \item Inference of optical DeiT model with different noise levels. 
    \item Estimate energy and latency of different DeiT/BERT models. We support DeiT-T/S/B and BERT-B/L, where you can adjust the sequence lengths for BERT. 
\end{itemize}

\subsection{Methodology}

Submission, reviewing and badging methodology:

\begin{itemize}
  \item \url{https://www.acm.org/publications/policies/artifact-review-badging}
  \item \url{http://cTuning.org/ae/submission-20201122.html}
  \item \url{http://cTuning.org/ae/reviewing-20201122.html}
  \item  \url{https://sc21.supercomputing.org/submit/reproducibility-initiative/ad-ae-appendix-process-badges/index.html#section16}
\end{itemize}